\documentclass[journal, draftclsnofoot, one column, 12pt]{IEEEtran}
\IEEEoverridecommandlockouts
 \pdfoutput=1
\usepackage{ushort}
\usepackage[mathcal]{euscript}
\usepackage{amsmath}
\usepackage{amsthm}
\usepackage{amssymb}
\usepackage{graphicx}
\usepackage{epstopdf}
\usepackage{booktabs}
\usepackage{colortbl}
\usepackage{amsmath}
\usepackage{xcolor}
\usepackage{graphicx}
\usepackage{algpseudocode}
\usepackage{algorithmicx}
\usepackage{algorithm}
\algnewcommand{\algorithmicgoto}{\textbf{go to}}%
\algnewcommand{\Goto}[1]{\algorithmicgoto~\ref{#1}}%
\usepackage{cite}
\usepackage{rotating}
\usepackage{array}
\usepackage{tensor}

\newtheorem{Thrm}{Theorem}

\newtheorem{Cor}{Corollary}
\newtheorem{Def}{Definition}

\def \te {\text{e}}
\def \t {\text}

\begin{document}

\title{A Delay Optimal MAC and Packet Scheduler for Heterogeneous M2M Uplink}

\author{\IEEEauthorblockN{Akshay Kumar, Ahmed Abdelhadi and Charles Clancy}\\
\IEEEauthorblockA{Hume Center, Virginia Tech\\
Email:\{akshay2, aabdelhadi, tcc\}@vt.edu}
\thanks{This research is based upon work supported by the National Science Foundation under Grant No. 1134843. 
}
}

\maketitle 

\vspace{-0.80cm}
\begin{abstract}
The uplink data arriving at the Machine-to-Machine (M2M) Application Server (AS) via M2M Aggregators (MAs) is fairly heterogeneous along several dimensions such as maximum tolerable packet delay, payload size and arrival rate, thus necessitating the design of Quality-of-Service (QoS) aware packet scheduler. In this paper, we classify the M2M uplink data into multiple QoS classes and use sigmoidal function to map the delay requirements of each class onto utility functions. We propose a proportionally fair delay-optimal multiclass packet scheduler at AS that maximizes a system utility metric. We note that the average class delay under any work-conserving scheduling policy can be realized by appropriately time-sharing between all possible preemptive priority policies. Therefore the optimal scheduler is determined using an iterative process to determine the optimal time-sharing between all priority scheduling policies, such that it results in maximum system utility. The proposed scheduler can be implemented online with reduced complexity due to the iterative optimization process. We then extend this work to determine jointly optimal MA-AS channel allocation and packet scheduling scheme at the MAs and AS. We first formulate a joint optimization problem that is solved centrally at the AS and then propose a low complexity distributed optimization problem solved independently at MAs and AS. We show that the distributed optimization solution converges quickly to the centralized optimization result with minimal information exchange overhead between MAs and AS. Using Monte-Carlo simulations, we verify the optimality of the proposed scheduler and show that it outperforms other state-of-the-art packet schedulers such as weighted round robin, max-weight scheduler etc. Another desirable feature of proposed scheduler is low delay jitter for delay-sensitive traffic. 
\end{abstract}

\begin{IEEEkeywords}
M2M, Latency, Quality-of-Service, MAC, MultiClass Scheduler, Convex Optimization
\end{IEEEkeywords}

\section{Introduction}
The Machine-to-Machine (M2M) market has witnessed a steady exponential growth over the past few years due to its plethora of both commercial and residential use-cases such as in smart healthcare, vehicle tracking, smart home, security and surveillance etc. \cite{M2Mgrowth}. As per Machina Research \cite{machinaResearch}, the total number of M2M connections globally will increase to $29$~billion by year 2024. Since M2M applications provides intelligent services by processing the sensory data to monitor and control the nodes, the M2M uplink is more demanding than the M2M downlink and thus attracted more attention from the M2M research community.

 The M2M uplink traffic unlike the human-originated traffic comprises of short-lived sessions involving transactions of a few hundred bytes. The M2M traffic exhibits heterogeneity in several dimensions such as maximum packet delay, packet size and arrival rate, depending on the use-case. For example, consider the M2M traffic generated by smart meters in a residential area \cite{smartGrid}. As per the UCA OpenSG specification (described in \cite{openSG}), at one extreme are the firmware and software updates with maximum latency of $5$~s, payload less than $50$~B and arrivals less than $0.5$~messages/day/device. On the other extreme are meter reading messages with high payload of around $1.2$~kB, maximum packet delay greater than $60$~s and $6$ messages/day/device. Due to the diverse characteristics of traffic generated from different M2M use-cases, the design of delay-efficient packet schedulers for M2M uplink is essential, particularly for delay-sensitive applications. The problem is exacerbated with increase in size of M2M network as the available computational and communication resources are shared among large number of machines. This makes it even more difficult to provide real-time service due to simultaneous access attempts from multiple sensors \cite{3GPPreport}.

\subsection{Motivation and Related Work}
Most of the existing M2M packet schedulers are designed for specific wireless technology such as LTE (see \cite{Gotsis12} and references therein). The schedulers proposed in these works use variants of Access Grant Time Interval algorithm \cite{Lein11} that allocates resources to the sensor nodes over a periodic time-interval in a fixed or dynamic manner. Nusrat et. al. in \cite{Afrin13} proposed a LTE uplink packet scheduler for M2M traffic with a goal of maximizing the fraction of packets that meet their service deadline. Ho and Huang in \cite{Ho12} proposed several energy-efficient resource allocation schemes for M2M traffic in LTE. However, it did not consider the delay requirements of machines in the design of schedulers, which is a more appropriate indicator of QoS of machines. Recently, a dynamic channel-, and QoS-aware heuristic was proposed for uplink LTE traffic \cite{Lioumpas11}. Another set of work focused on delay-efficient packet scheduling heuristics for M2M uplink that maps delay requirement of periodic and aperiodic traffic onto service utility functions and then seeks to maximize a proportionally-fair system utility metric \cite{kumarHeuristicSyscon, kumarHeuristicMilcom}. Unlike our present work, these works propose heuristic schedulers and/or designed for specific Medium Access Control architecture of a particular wireless communication standard. 

Another line of work considers the design of packet schedulers for real-time control applications (see \cite{Buttazzo11} and references therein) and assume hybrid task sets comprising of periodic tasks with hard deadlines and aperiodic tasks with hard, soft or non-real-time requirements. The goal of these schedulers is to guarantee schedulability for critical tasks (with hard deadlines) while providing a best-effort service for other tasks. The periodic tasks from different flows are processed by assuming a fixed-priority assignment and with service deadline equal to the arrival period of that flow, while the aperiodic tasks from various flows are aggregated and served using First-Come-First-Serve policy. Some of the popular algorithms include Rate Monotonic scheduling \cite{LiuLayland73}, Polling Server, Deferrable Server \cite{Strosnider95}, Priority Exchange \cite{Lehoczky87}, Sporadic Server \cite{Sprunt89} and Slack-Stealing \cite{Lehoczky92}.  However for most of the practical M2M systems, the arrival rate and delay requirements are potentially different for each aperiodic flow and thus aggregating the aperiodic flows will degrade the overall delay performance. Furthermore, these schemes define a \emph{server}, a virtual periodic task with certain computation time and priority level to process outstanding aperiodic tasks. But the determination of delay-optimal \emph{server} period and \emph{server} capacity for given traffic characteristics is not clear. 

 Besides these works, a number of state-of-the-art packet schedulers have been designed for general queuing networks. Fair queuing \cite{Demers89, Greenberg90} provides perfect fairness among different traffic classes but almost all its implementations suffer from high operational complexity and also do not consider the delay-heterogeneity of traffic, which is an important attribute of M2M uplink traffic. To get around these issues, weighted round robin (WRR), deficit round robin (DRR) and weighted fair scheduling (WFS) have been proposed \cite{Nagle87, Parkeh93, Shreedhar96}, that can guarantee proportionally fair service by setting the weights in accordance with packet size and delay requirements of different flows. However, the determination of the delay-optimal weights for these schemes is not trivial and are usually assigned by the site administrator based on certain criteria. Tassiulas et. al. \cite{Tassiulas} proposed the throughput-optimal max-weight scheduling algorithm, but it may result in highly unfair resource allocation when the arrival rates for different flows are skewed.
 
In this work, we consider a generic M2M network wherein the uplink data from a local group of sensors is first aggregated at a M2M aggregators (MA) and then the data from multiple such aggregators is finally processed at an Application Server (AS) residing in the M2M core network. The uplink sensor traffic is classified into multiple Quality-of-Service (QoS) classes based on the packet arrival rate, payload size and maximum packet delay requirements of different applications. We address the shortcomings of the existing packet schedulers by proposing a low-complexity, proportionally-fair delay-optimal multiclass packet scheduler. It also results in a much lower delay jitter for the delay-sensitive traffic which is highly desirable. But this is achieved at the expense of higher delay jitter for delay-tolerant traffic which is usually not a big concern due to its delay-tolerant nature. More details of the proposed packet scheduler along with the other contributions of this paper are provided next. 
\subsection{Contributions and Outcomes}
\emph{Proportionally-fair, delay-optimal packet scheduler at AS:} We first design a delay-optimal multiclass packet scheduler at the AS, while neglecting other network delays. We use sigmoidal function to represent the utility of certain service delay experienced by packets of a class and then aggregate the utility values for all the classes into a proportionally-fair system utility metric. We then define a delay-optimal scheduler as the one that maximizes the proposed system utility metric. We note that the average delay of each QoS class under any work-conserving\footnote{A work-conserving scheduler does not result in server being idle while there are jobs in the queue waiting for service.} scheduling policy can be realized by appropriately time-sharing between all possible preemptive\footnote{Hereafter, we drop the qualifier 'preemptive' for succinctness.} priority scheduling policies. Therefore, we determine the optimal scheduler by solving for the optimal fraction of time-sharing between different priority scheduling policies that maximizes the system utility.

\emph{Low-complexity optimal scheduler:} We significantly reduce the computational complexity of determining the optimal scheduler by reformulating and solving the original single-stage optimization problem as an iterative optimization algorithm. This reduction in complexity is achieved without any loss in delay-optimality of the proposed scheduler. 

\emph{Joint channel assignment and packet scheduler at MAs and AS:} We extend our work to the joint problem of channel assignment for MA-AS links and packet scheduler at MAs and AS. To solve it, we first formulate and solve a joint single-stage optimization problem to be solved centrally at AS. We also propose an alternative low-complexity, distributed and iterative optimization approach. 
 
\emph{Superior delay performance of proposed scheduler:} Using Monte-Carlo simulations, we verify the correctness of the analytical result for proposed optimal scheduler at AS and show that it outperforms other state-of-the-art schedulers such as WRR, max-weight scheduler, WFS and fixed priority scheduling.  It also results in a much lower delay jitter for the delay-sensitive traffic which is highly desirable. But this is achieved at the expense of higher delay jitter for delay-tolerant traffic which is usually not a big concern due to its delay-tolerant nature.  We then show that the centralized joint MA-AS packet scheduler has superior delay-performance as compared to other schedulers and the distributed optimization scheduler results in optimal (near-optimal) delay-performance for exponential (constant) service distribution. 

\emph{Flexible M2M scheduler:} The proposed M2M packet scheduler is agnostic to the communication standard and the hardware-software architecture used for M2M network. It can also adapt to time-varying characteristics of M2M traffic by either solving the optimization problem on-the-fly or looking up for the optimal scheduler using a Look-Up-Table which can be populated during a training phase prior to the deployment of the proposed scheduler.  

The rest of the paper is organized as follows. Section~\ref{sysModel_M2M2} introduces the system model for a generic M2M uplink. Then in Section~\ref{probForm_M2M2}, we define the utility functions for each class, formulate both the single-stage and iterative utility maximization problem. We prove the iterative optimization problem is convex, and solve it to obtain the optimal scheduler in Section \ref{optSch}. We then present the centralized and distributed optimization approach in Section~\ref{jointOptimal} for finding the jointly optimal scheduler at MAs and AS. Then in Section~\ref{jointSubcarrierScheduler}, we study the problem of joint channel assignment for MA-AS links and packet scheduling at MAs and AS. Section~\ref{Results_M2M2} presents the simulation results. Finally Section~\ref{concl} draws some conclusions.

\section{System Model}
\label{sysModel_M2M2}
\begin{figure*}%
\centering
\includegraphics[height = 2.40 in, width = 3.9 in]{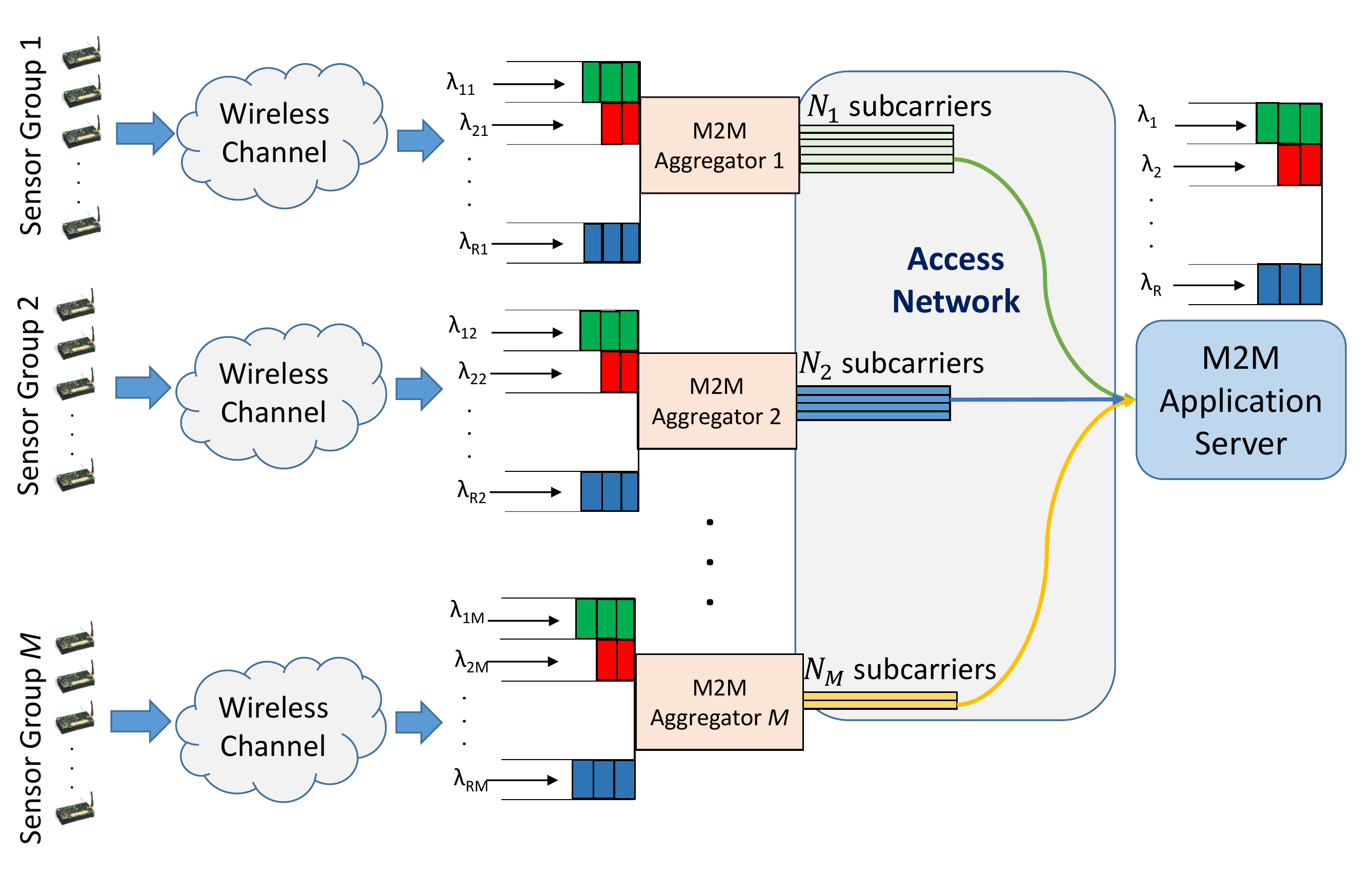}%
\caption{System model for a heterogeneous M2M uplink system showing the queuing process at M2M aggregators and the M2M application server. The different packet colors at AS indicate different delay classes.}%
\label{systemModel_M2M2}%
\vspace{13pt}
\end{figure*}

Fig.~\ref{systemModel_M2M2} shows the system model for a general M2M uplink system. The data transmitted from each local group of sensors is aggregated at a M2M aggregator (MA) and classified into $R$ classes based on the packet size and packet delay requirements. We assume that the packet arrival process for class $i$ at $k^{\t{th}}$ MA is Poisson with rate $\lambda_{ik}$ \cite{Dhillon14}. The data from the set of $M$ MAs is concurrently transmitted to the M2M Application Server (AS) using a set of orthogonal channels\footnote{For instance, this can be achieved using OFDMA across the MAs to transmit the buffered packets to AS}. This avoids packet collisions between different MA-AS links due to simultaneous transmission, without significantly increasing the complexity at MAs or AS. Furthermore, since the M2M traffic from multiple sensors is aggregated at each MA, assigning dedicated resources for each MA-AS link would not result in any significant resource wastage.

 We assume that the total spectrum available for the MA-AS communication is divided into $N$ orthogonal subcarriers, each of bandwidth $W$. The link between $k^{\t{th}}$ MA and AS is allocated $N_k$ subcarriers, depending on the incoming load and traffic delay-requirements at the MA. We assume that each MA transmits at a nominal power to achieve a certain  subcarrier signal-to-noise ratio (SNR) $\eta$ at the AS under Additive White Gaussian Noise (AWGN) channel conditions. The channel between the MAs and AS is assumed to be independent slow Rayleigh fading since the MAs and AS are stationary nodes. Furthermore, we assume flat fading because of the small bandwidth requirement for the M2M system relative to the microwave carrier frequencies. Then the resultant capacity per subcarrier at $k^{\t{th}}$ MA, $C_{s,k}$, is given by Shannon's capacity formula as,
\begin{equation}
C_{s,k} = W {\t{log}}_2(1+{\|h_{k}\|}^2 \eta)~\t{bits/sec},
\end{equation}
where $h_k$ is the channel gain at a given instant between the $k^{\t{th}}$ MA and the AS.
The total capacity for $k^{\t{th}}$ MA is simply $C_k = N_k C_{s,k}$ bits/sec. 
 
The packets of each class $i$ from the $M$ MAs are then aggregated at AS resulting in a Poisson process for class $i$ with rate $\lambda_i = \sum_{k=1}^M \lambda_{ik}$. The packet size may vary for different applications but we assume that the packet size for all the M2M applications mapped to the same class is approximately same. Therefore, the packet size for class $i$ is modeled as a constant of $s_i$ bits. Then the packet transmission rate for class $i$ at $k^{\t{th}}$ MA is $\mu_{ik} = C_k/s_i$ packets/sec. Let $C_o$ denote the service rate at AS, then the the service rate for class $i$ at AS is $\mu_i = C_o/s_i$ packets/sec. 

The total time spent by a packet in the system consists of transmission time at sensor and MA, propagation and congestion delay for the sensor-MA link and MA-AS link, queuing delay at MA and AS and lastly, the service time at AS. The  propagation delay is usually negligible compared to other delay components and can be safely ignored. The packet transmission time at sensor is also negligible due to small packet size relative to the channel capacity. Furthermore, due to low data rate for traffic at each sensor and small number of sensors in each group, the congestion delay for the sensor-MA link can also be ignored. However, due to traffic aggregation at each MA, the MA-AS link is usually resource constrained. Therefore, we need to account for queuing and service delay at both MA and AS. Since, we assume dedicated orthogonal channels for the MA-AS link, we ignore the congestion delay for the MA-AS link.

We first study the basic packet scheduling problem at AS accounting only for the queuing and service delay at AS. We then extend this analysis to a joint packet scheduler for MAs and AS in Section~\ref{jointOptimal}. Lastly in Section~\ref{jointSubcarrierScheduler}, we study the problem of joint subcarrier assignment and packet scheduling at MAs and AS. 	 	

\section{Problem Formulation}
\label{probForm_M2M2}
We first use a generic sigmoidal function \cite{AbdelhadiCNC2014, AbdelhadiPIMRC2013} to map the latency\footnote{We use the terms \lq delay\rq~and \lq latency\rq~interchangeably. Both refer to the sum of queuing delay and service time.} requirements of $i^{\t{th}}$ class, $l_i$, onto a utility function as, 
\begin{align}
U_i(l_i) = 1- c_i\left(\frac{1}{1+{\text{e}}^{-a_i(l_i-b_i)}}-d_i \right) \label{utilClassi}
\end{align}
where, $c_i=\frac{1+{\text{e}}^{a_ib_i}}{{\text{e}^{a_ib_i}}}$ and $d_i = \frac{1}{1+{\text{e}^{a_ib_i}}}$. Note that $U_i(0)=1$ and $U_i(\infty) = 0$. The parameter $a_i$ is the utility roll-off factor and the inflection point for $U_i$ occurs at $l_i = b_i$.

\begin{figure}
\centering
\includegraphics[scale=0.45]{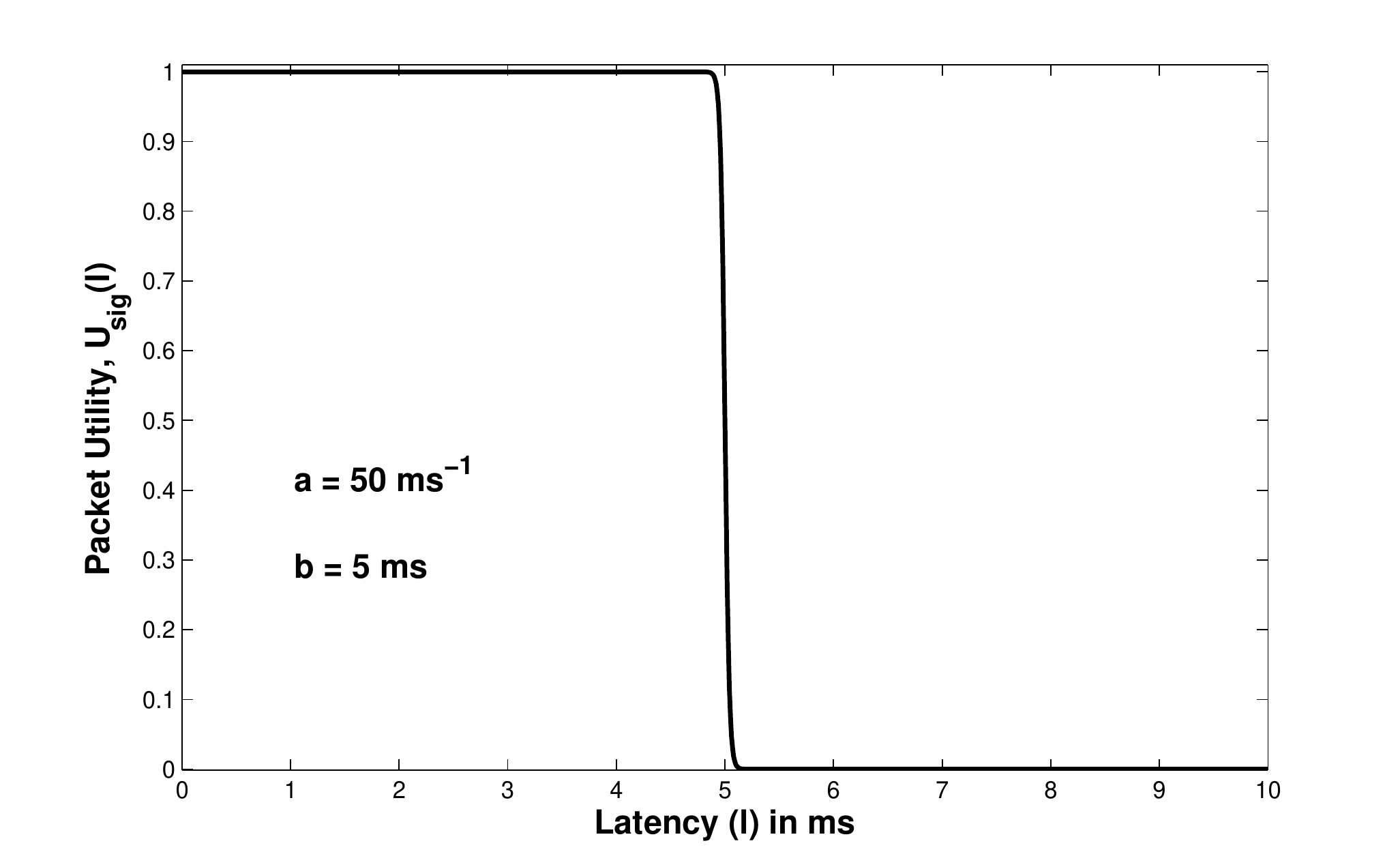}   
\caption{Sigmoidal utility function for delay-sensitive traffic.}
\label{puUtilFcn} 
\end{figure}

\begin{figure}
\centering{
\includegraphics[scale=0.45]{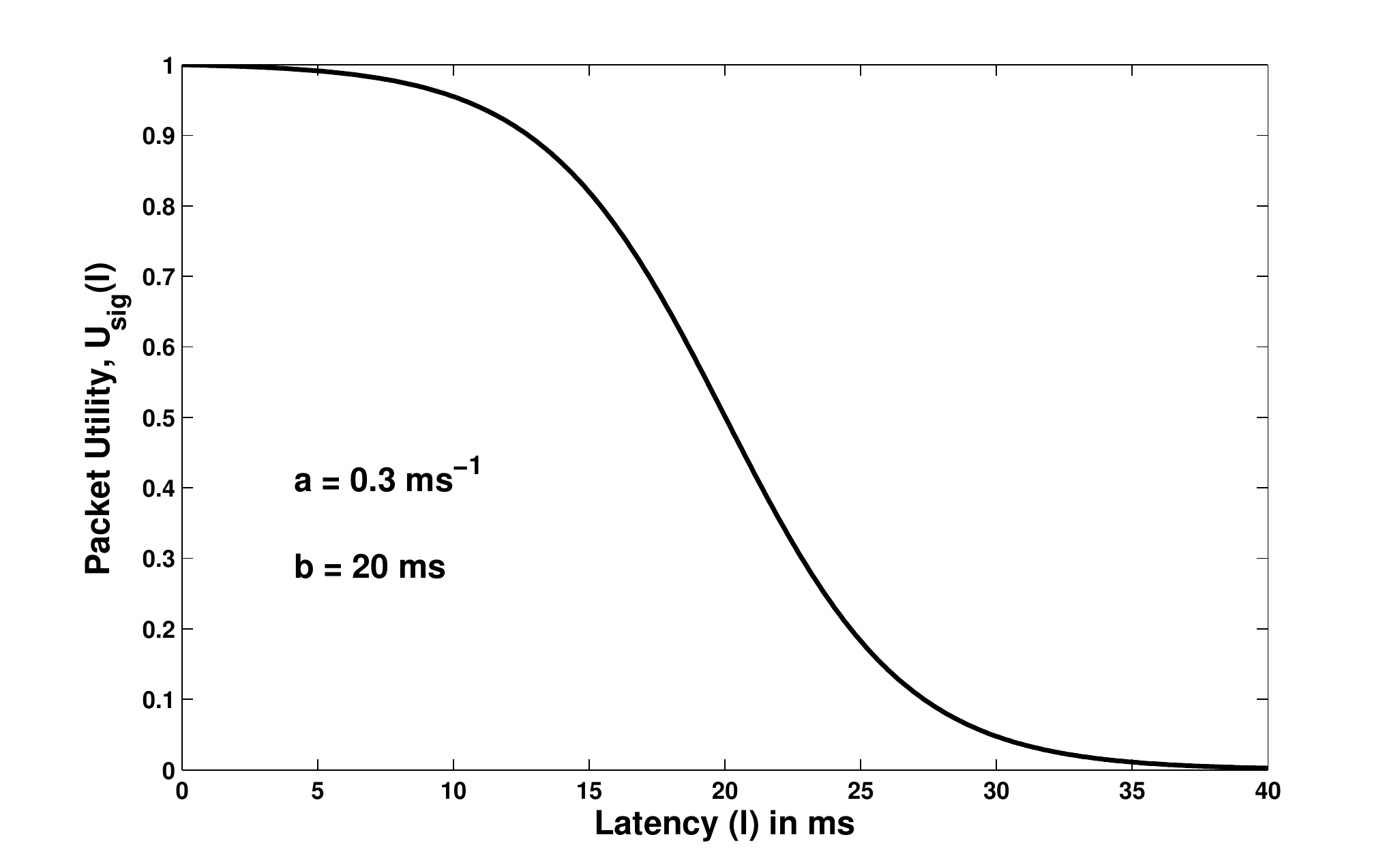}%
\caption{Sigmoidal utility function for delay-tolerant traffic.}}
\label{edUtilFcn} 
\end{figure}

The sigmoidal function is versatile to represent diverse delay requirements by appropriately choosing the parameters $a$ and $b$. For high $a$ and low $b$, the utility function becomes \lq brick-walled\rq~(see Fig.~\ref{puUtilFcn}) and is a good fit for delay-sensitive applications. On the other hand, at low $a$ and high $b$, the sigmoidal utility function is a good fit for delay-tolerant applications as shown in Fig.~\ref{edUtilFcn}. 

\subsection{System utility function} For a given scheduling policy $\mathcal{P}$, we define a proportionally fair system utility function as,
\begin{equation}
V(\mathcal{P}) = \prod_{i=1}^R \mathbf{U_i}^{\beta_{i}}(\mathcal{P}),
\label{sysUtil_M2M2}
\end{equation}
where $\mathbf{U_i}(\mathcal{P})$ is the average utility of traffic of $i^{\t{th}}$ class in the steady state given as,
\begin{align}
\mathbf{U_i}(\mathcal{P}) &= U_i \left(\lim_{{T_{\text{s}}} \rightarrow \infty} \frac{\sum_{j=1}^{M_{i}(T_{\text{s}})} l_i^j(\mathcal{P})}{M_{i}(T_{\text{s}})}\right), 
\end{align}
where $M_{i}(T_{\text{s}})$ is the number of packets of class $i$ served in time $T_{\text{s}}$ and $l_i^j$ is the latency of the $j^{\t{th}}$ packet of $i^{\t{th}}$ class. The parameters $\beta_{i}$ indicates the relative importance of utility of $i^{\t{th}}$ class towards the system utility.

\subsection{Single-stage Optimization Problem}
\label{directOptimProb}
We now formulate a single-stage optimization (\textbf{SSO}) problem to solve for the optimal scheduling policy. We first define the sets $\mathsf{R}=\{1, 2, \cdots, R!\}$ and $\mathcal{R} = \{1, 2, \cdots, R\}$. Assume that in a sufficiently large time interval\footnote{We assume the time interval under consideration is large enough to observe steady state queuing behavior.}, the AS serves the buffered traffic using $j^{\t{th}}$ priority order\footnote{There are $R!$ different priority orders for a set of $R$ classes. For instance if $R=3$, the different priority orders are $123, 132, 213, 231, 312, 321$, where $pqr$ denotes that class $p$ has higher priority than class $q$ which in turn is greater than priority of class $r$.} for $\gamma_j$ fraction of the time $\forall~j\in \mathsf{R}$. Then we can write the following equation for the average latency $l_i$ of class $i$,
\begin{equation}
\label{latEqSet_direct}
l_i = \sum_{j=1}^{R!} \gamma_j l_{i,j}^s~\forall~i \in \mathcal{R},
\end{equation}
where $l_{i,j}^s$ represents the average latency of class $i$ using $j^{\t{th}}$ priority order in the direct optimization problem.

Let $\mathcal{H}_{i,j}$ denote the set of classes with higher priority than class $i$ in the $j^{\t{th}}$ priority order. Then using the existing results on latency of M/G/1 preemptive priority queuing systems by Bertsekas et. al. \cite{gallager}, we get,
\begin{align}
l_{i,j}^s &= \frac{D_{\mathcal{H}_{i,j},i} + (\rho_{\mathcal{H}_{i,j}}-\rho_{i})(1/\mu_{i})}{\rho_{\mathcal{H}_{i,j}}(\rho_{\mathcal{H}_{i,j}}-\rho_i)} ~ \forall~i, \label{lat_i_direct}
\end{align}
where $\rho_i = \lambda_i/\mu_i$ is the AS utilization factor for class $i$ and $\rho_{\mathcal{H}_{i,j}}= 1-\sum_{m\in \mathcal{H}_{i,j}} \rho_m$. Since each class is a M/D/1 queue, the residual time $D_{\mathcal{H}_{i,j},i}$ is given by,
\begin{equation}
D_{\mathcal{H}_{i,j},i} = \sum_{k\in \mathcal{H}_{i,j} \cup \{i\}}\frac{\rho_{k}}{2\mu_{k}}.                                         
\label{resTime_direct}
\end{equation}

Since this scheduling policy is characterized by $\gamma = \{\gamma_j: j \in \mathsf{R}\}$, the system utility in \eqref{sysUtil_M2M2} becomes,
\begin{equation}
V(\gamma) = \prod_{i=1}^{R} \bold{U_i}^{\beta_{i}}(\gamma ).
\label{systUtil1}
\end{equation}

Now the optimal scheduling policy is determined by solving for $\gamma$ that maximizes $V(\gamma)$. Exploiting the strictly increasing nature of logarithms, we get the following utility-maximization problem,
\begin{equation}
\begin{aligned}
\textbf{SSO:} &\operatorname*{max}_{\gamma} & & \text{log}\left(V(\gamma)\right) = \sum_{i=1}^R {\beta_i}\text{log}\left(U_i(\gamma)\right)  \\
&\text{s.t.} & &\gamma_j \geq 0 ~\forall~j\in\mathsf{R},  \\
& & & \sum_{j=1}^{R!} \gamma_j = 1.
\end{aligned}
\label{optimProb1}
\end{equation}
We note that the optimization is done over $R!$ variables and thus its complexity is enormous even at moderate values of $R$. For instance, for $R=7$ the number of variables is $5040$.

\subsection{Iterative Optimization Problem}
\label{iterativeOptim}
To reduce the complexity of the above optimization problem, we reformulate it as an iterative optimization (\textbf{IO}) problem. Let $\mathcal{A}_r$ be set of any $r~(2\leq r\leq R)$ classes and $\mathcal{A}_r^{\mathsf{c}} = \mathcal{R}\backslash \mathcal{A}_r$, where $A\backslash B$ denotes the set difference $A-B$. We assume that each class in $\mathcal{A}_r^{\mathsf{c}}$ is assigned higher priority than classes in $\mathcal{A}_r$. If AS serves class $i \in \mathcal{A}_r$ with highest priority in $\mathcal{A}_r$ for $\alpha_i$ fraction of the time, then the average latency $l_i$ of class $i$ is given by,
\begin{equation}
\label{latEqSet}
l_i =  \alpha_i l_{i,i} + \sum_{j \in \mathcal{A}_r, j\neq i} \alpha_j l_{i,j}^* ~\forall~i\in\mathcal{A}_r,
\end{equation}
where $l_{i,i}$ denotes the latency of the class $i$ when it is served with highest priority. $l_{i,j}^{*}$ denotes the optimal latency for the class $i$ obtained on solving the $r-1$ class system with $\mathcal{A}_{r-1} = \mathcal{A}_{r}\backslash \{j\}$. Let us define $l_{\mathcal{A}_r} = \{l_i : i\in \mathcal{A}_r\}$ and $l_{\mathcal{A}_r \backslash \{j\}}^* = \{l_{i,j}^{*}: i \in \mathcal{A}_r, i\neq j\}$.  

Now reapplying the result for latency of M/D/1 preemptive priority queuing systems similar to \eqref{lat_i_direct}, we get,
\begin{align}
l_{i,i} &= \frac{D_{\mathcal{A}_r^{\mathsf{c}},i} + (\rho_{\mathcal{A}_r^{\mathsf{c}}}-\rho_{i})(1/\mu_{i})}{\rho_{\mathcal{A}_r^{\mathsf{c}}}(\rho_{\mathcal{A}_r^{\mathsf{c}}}-\rho_i)}~\forall~i \in \mathcal{A}_r. \label{lat_i_highest}
\end{align}
 
For the special case of $r=2$ in \eqref{latEqSet}, we note that $l_{i,j}^*$ is equivalent to a $l_{i,i}$ in $r=1$ system with $\mathcal{A}_r = \{i\}$. Hence, we obtain $l_{i,j}^*$ explicitly using \eqref{lat_i_highest} as,
\begin{equation}
\label{latRis2}
l_{i,j}^* = \frac{D_{\mathcal{A}_r^{\mathsf{c}},i} + \rho_j/\mu_j +  \rho_{\mathcal{R}}/\mu_{i}}{\rho_{\mathcal{R}}(\rho_{\mathcal{R}}+\rho_{i})}, \\
\end{equation}
Since this scheduling policy is characterized by $\alpha_{\mathcal{A}_r} = \{\alpha_i: i\in \mathcal{A}_r\}$, the system utility in \eqref{sysUtil_M2M2} becomes,
\begin{equation}
V(\alpha_{\mathcal{A}_r}) = \prod_{i \in \mathcal{A}_r} \bold{U_{i}}^{\beta_{i}}(\alpha_{\mathcal{A}_r}).
\label{systUtil2}
\end{equation}

Now the optimal scheduling policy for a $r$ class system with given $\mathcal{A}_r$, is determined by solving for $\alpha_{\mathcal{A}_r}$ that maximizes $V(\alpha_{\mathcal{A}_r})$. Exploiting the strictly increasing nature of logarithms, we obtain the following $r$-class optimization subproblem,
\begin{equation}
\begin{aligned}
\textbf{IO:} &\operatorname*{max}_{\alpha_{\mathcal{A}_r}} & & \text{log}\left(V(\alpha_{\mathcal{A}_r})\right) = \sum_{i \in \mathcal{A}_r} {\beta_{i}}~\text{log}\left(U_{i}(\alpha_{\mathcal{A}_r})\right)  \\
&\text{s.t.} & & \alpha_{i} \geq 0 ~\forall~i \in \mathcal{A}_r,  \\
& & & \sum_{i \in \mathcal{A}_r} \alpha_{i} = 1.
\end{aligned}
\label{optimProb2}
\end{equation}

\begin{Thrm}
\label{Theorem1}
The iterative optimization problem $\textbf{IO}$ in \eqref{optimProb2} is convex.
\end{Thrm}
\begin{IEEEproof}
 The proof outline begins with proving that $f^i(\alpha_{\mathcal{A}_r}) = {\beta_i}\text{log}\left(U_i(\alpha_{\mathcal{A}_r})\right)$ is concave $\forall~i \in \mathcal{A}_r$ and for all possible $\mathcal{A}_r$. Now the Hessian matrix of $f^i$ at the point $\alpha_{\mathcal{A}_r}$ is defined as,
\[ H^i(\alpha_{\mathcal{A}_r}) = \left( \begin{array}{cccc}
f^i_{11}(\alpha_{\mathcal{A}_r}) & f^i_{12}(\alpha_{\mathcal{A}_r}) & \cdots & f^i_{1r}(\alpha_{\mathcal{A}_r})\\
f^i_{21}(\alpha_{\mathcal{A}_r}) & f^i_{22}(\alpha_{\mathcal{A}_r}) & \cdots & f^i_{2r}(\alpha_{\mathcal{A}_r}) \\
         \vdots 					 &         \vdots             & \ddots &  \vdots									 \\
f^i_{r1}(\alpha_{\mathcal{A}_r}) & f^i_{r2}(\alpha_{\mathcal{A}_r}) & \cdots & f^i_{rr}(\alpha_{\mathcal{A}_r}) 
\end{array} \right),\]
where $f^i_{jk} = \frac{\partial^2 f^i}{\partial \alpha_j \partial \alpha_k}$. 

On solving, we get the following result,
\begin{equation}
f^i_{jk}(\alpha_{\mathcal{A}_r})  = \frac{-\beta_i \theta_i a_i^2 \zeta_{i,j} \zeta_{i,k}}{{(1+\theta_i)}^2},
\end{equation}
where $\zeta_{i,j} = l_{i,j}^*$ when $i \neq j$ and is $l_{i,i}$ for $i=j$. We define $\theta_i = \te^{-a_i(l_i-b_i)}$.

Therefore, $H^i(\alpha_{\mathcal{A}_r})$ is given by,  $H^i(\alpha_{\mathcal{A}_r}) =$
\begin{equation}
\frac{-\beta_i \theta_i a_i^2}{{(1+\theta_i)}^2} \!\left( \begin{array}{cccc}
\zeta_{i,(1)}^2 & \zeta_{i,(1)} \zeta_{i,(2)} & \cdots & \zeta_{i,(1)} \zeta_{i,(r)} \\
\zeta_{i,(1)} \zeta_{i,(2)} & \zeta_{i,(2)}^2 & \cdots & \zeta_{i,(2)} \zeta_{i,(r)} \\
         \vdots 					 &         \vdots             & \ddots &  \vdots									 \\
\zeta_{i,(1)} \zeta_{i,(r)} & \zeta_{i,(2)} \zeta_{i,(r)} & \cdots & \zeta_{i,(r)}^2
\end{array} \right),
\label{hessian}
\end{equation}
where for ease of representation, we defined the shorthand notation $(k) = \mathcal{A}_r(k)$, i.e., the $k^{\t{th}}$ element of set $\mathcal{A}_r$. 
Now to prove that $f^i$ is a concave function, it is sufficient to prove that $H^i$ is a Negative Semi-Definite (NSD) matrix for all $\alpha_{\mathcal{A}_r}$ that satisfies the constraints in \eqref{optimProb2}. Let $\Delta_k$ denote a $k^{\t{th}}$ order principal minor of $H^i$. Then $H^i$ is NSD if and only if $(-1)^k \Delta_k \geq 0$ for all principal minors of order $k = 1, 2, \cdots, r$.

From ~\eqref{hessian}, we obtain the following, $\Delta_1 =$
\begin{equation}
\begin{aligned}
& \frac{-\beta_i \theta_i a_i^2 \zeta_{i,(1)}^2}{{(1+\theta_i)}^2}, \frac{-\beta_i \theta_i a_i^2 \zeta_{i,(2)}^2}{{(1+\theta_i)}^2}, \cdots, \frac{-\beta_i \theta_i a_i^2 \zeta_{i,(r)}^2}{{(1+\theta_i)}^2},\\
&\Delta_2 = 0, \Delta_3 = 0, \cdots, \Delta_r = 0.
\end{aligned}
 \end{equation}
Clearly all $\Delta_1 < 0$ and all higher order principal minors are $0$. Therefore $H^i(\alpha_{\mathcal{A}_r})$ is NSD for all $\alpha_{\mathcal{A}_r}$. Hence, $f^i(\alpha_{\mathcal{A}_r})$ is concave for all $i \in \mathcal{A}_r$. This implies that the aggregated utility natural logarithm $\text{log}\left(V({\alpha_{\mathcal{A}_r}})\right) = \sum_{i=1}^r f^i(\alpha_{\mathcal{A}_r})$ is concave.

Therefore, the optimization problem in \eqref{optimProb2} has concave objective function with affine constraints. Hence, the optimization problem in \eqref{optimProb2} is convex.
\end{IEEEproof}

\begin{Cor}
\label{Corollary1}
The direct optimization problem $\textbf{SSO}$ in \eqref{optimProb1} is also convex.
\end{Cor}
\begin{IEEEproof}: We note that the problem $\textbf{SSO}$ in \eqref{optimProb1} is a special case of the problem $\textbf{IO}$ in \eqref{optimProb2}. Specifically, the set $\mathcal{A}_r$ in $\textbf{IO}$ is equal to $\mathcal{R}$ in $\textbf{SSO}$. Further, each latency equation in \eqref{latEqSet_direct} is a function of $R!$ optimization variables $\gamma$, whereas in \eqref{latEqSet} it is a function of $r$ variables $\alpha_{\mathcal{A}_r}$. We can consider this as $\alpha_{\mathcal{A}_r}$, $l_{i,j}^*$ and $l_{i,i}$ in $\textbf{IO}$ being replaced by $\gamma$, $l_{i,j}^s$ and $l_{i,i}^s$ in $\textbf{SSO}$ respectively. Hence, the problem $\textbf{SSO}$ is just a special case of $\textbf{IO}$ with same structure. 

Now from Theorem~\ref{Theorem1}, we note that the optimization problem in \eqref{optimProb2} is convex. Therefore, we can conclude that the direct optimization problem $\textbf{SSO}$ in \eqref{optimProb1} is also convex. 
\end{IEEEproof}

\section{Optimal Scheduler}
\label{optSch}
We now solve the dual problem of \eqref{optimProb2} to determine the optimal solution. We first define the Lagrangian as,
\begin{align}
L({\alpha_{\mathcal{A}_r}},\eta) =  \sum_{i \in \mathcal{A}_r} {\beta_i}\text{log}\left(U_i({\alpha_{\mathcal{A}_r}})\right) - \eta\left(\sum_{i \in \mathcal{A}_r} \alpha_i - 1\right),
\end{align}
where $\eta \geq 0$ is a Lagrange multiplier.
The dual problem is formulated as follows,
\begin{equation}
\begin{aligned}
&\operatorname*{min}_{\eta} \operatorname*{max}_{{\alpha_{\mathcal{A}_r}}} L({\alpha_{\mathcal{A}_r}},\eta)\\
&\text{s.t.} \quad \eta \geq 0, \alpha_i \geq 0,~ \forall~i \in \mathcal{A}_r.
\end{aligned}
\end{equation}

At the optimal solution $({\alpha_{\mathcal{A}_r}^*},\eta^*)$, we have,
\begin{align} 
&\frac{\partial}{\partial \alpha_i} \left[ \text{log}(V({\alpha_{\mathcal{A}_r}})) - \eta\left(\sum_{j \in \mathcal{A}_r} \alpha_j - 1\right)\right]_{{\alpha_{\mathcal{A}_r}^*},\eta^*} = 0, \label{maxUtil} \\
& \sum_{j \in \mathcal{A}_r} \frac{\beta_j}{U_j({\alpha_{\mathcal{A}_r}})} \frac{\partial}{\partial \alpha_i} U_j({\alpha_{\mathcal{A}_r}}) \Big|_{{\alpha_{\mathcal{A}_r}} = {\alpha_{\mathcal{A}_r}^{*}}} = \eta^*~\forall~i. \label{derivativeSysUtil}
\end{align}

Using \eqref{utilClassi} and \eqref{latEqSet}, we have,
\begin{align}
\frac{\partial}{\partial \alpha_i} U_j({\alpha_{\mathcal{A}_r}}) &= \frac{-c_j \beta_j a_j \te^{-a_j(l_j-b_j)} \zeta_{j,i}}{{\left(1+ \te^{-a_j(l_j-b_j)}\right)}^2}~ \forall~i. 
\label{derivUtil} 
\end{align}

Substituting \eqref{derivUtil} in \eqref{derivativeSysUtil}, followed by some algebraic manipulations, we get,
\begin{align}
\sum_{j\in \mathcal{A}_r} \frac{k_{j,i}} {1+ \te^{-a_j(l_j-b_j)}} \Big|_{{\alpha_{\mathcal{A}_r}} = {\alpha_{\mathcal{A}_r}^{*}}} + \eta^* = 0 ~ \forall~i.
\label{finalSoln}
\end{align}
where $k_{j,i} = \beta_j a_j \zeta_{j,i}$ is a constant. 

Also, at the optimal solution $({\alpha_{\mathcal{A}_r}^*},\eta^*)$, we have,
\begin{align} 
&\frac{\partial}{\partial \eta} \left[ \text{log}(V({\alpha_{\mathcal{A}_r}})) - \eta\left(\sum_{j\in \mathcal{A}_r} \alpha_j - 1\right)\right]_{{\alpha_{\mathcal{A}_r}^*},\eta^*} = 0,   \\
&\implies \sum_{j\in \mathcal{A}_r} \alpha_j^* = 1.\label{derivativeSysUtil1}
\end{align}

We determine the optimal solution by solving for $(\alpha_{\mathcal{A}_r}^*, \eta^*)$ using \eqref{finalSoln} and \eqref{derivativeSysUtil1} simultaneously. The constraint $\alpha_i^* \geq 0$ is enforced by setting $\alpha_i^* = 0$ if it is negative and then we solve again for $\alpha_{\mathcal{A}_r}^*$. Now the optimal latency for $r$ class system, $l_{\mathcal{A}_r}^*$, is obtained by setting $\alpha_{\mathcal{A}_r} = \alpha_{\mathcal{A}_r}^*$ in \eqref{latEqSet}.

The iterative algorithm for determining the optimal multiclass scheduler is described in Algorithm~\ref{propScheduler}. The function \textsc{optSch} is initialized with $\mathcal{A}_r = \mathcal{R}$, i.e., $r=R$.  At the $r$-class iteration, we iterate through each class $i \in \mathcal{A}_r$, assign it as highest priority in $\mathcal{A}_r$ which is equivalent to moving it from $\mathcal{A}_r$ to $\mathcal{A}_r^\mathsf{c}$. We then determine the optimal scheduler for resultant $r-1$ class system, $\mathcal{A}_{r-1} = \mathcal{A}_r\backslash \{i\}$. This downward recursion is performed until $r=2$ for which the optimal solution is determined using procedure outlined in Section~\ref{optSch}. Then the optimal solution from $r-1$ class system, $(l_{\mathcal{A}_{r-1}}^*,\alpha_{\mathcal{A}_{r-1}}^*)$ is returned upwards to to determine the optimal scheduler for $r$ class system, $\mathcal{A}_{r}$. Finally, the $\alpha_{\mathcal{A}_r}^{*} \forall r, \mathcal{A}_r$ are aggregated to determine the fraction of time the scheduler at AS uses each of the $R!$ priority orders. We denote this by $\gamma = \{\gamma_i: i\in\mathsf{R}\}$, as described in Section~\ref{directOptimProb}.

\begin{algorithm}[!h]
\caption{Proposed Iterative $r$-Class Scheduler}
\label{propScheduler}
\small{
\begin{algorithmic} 
	\Function{optSch}{$\mathcal{A}_r$}
		\If{$\text{length}(\mathcal{A}_r)>2$} 
			\For{$i \in \mathcal{A}_r$}		
			  \If{$i < \text{max}(\mathcal{A}_r^\mathsf{c})$}
			     \State Reuse $l_{\mathcal{A}_r\backslash \{i\}}^*$ and $\alpha_{\mathcal{A}_r \backslash \{i\}}^*$. 
			  \Else
			     \State $(l_{\mathcal{A}_r\backslash \{i\}}^*,\alpha_{\mathcal{A}_r \backslash \{i\}}^*)$ = \textsc{optSch}$(\mathcal{A}_r\backslash \{i\})$ 		  
			  \EndIf
              \State Determine $l_{i,i}$ using Eq~\eqref{lat_i_highest}.						
			\EndFor	
		\Else \Comment{For $r=2$}
				\State Use \eqref{lat_i_highest} and~\eqref{latRis2} for $l_{i,i}$, $l_{i,j}^*$.								
		\EndIf			
				\State Use method in Section~\ref{optSch} to solve for $l_{\mathcal{A}_r}^*$ and $\alpha_{\mathcal{A}_r}^*$. \\
				\Return $(l_{\mathcal{A}_r}^*,\alpha_{\mathcal{A}_r}^*)$.
		\EndFunction
	\end{algorithmic}	
	}
\end{algorithm}

\subsection{Complexity Analysis}
We now compare the complexity of solving the single-stage and iterative optimization problems.
\subsubsection{Single-stage optimization} For a $R$ class system, there are $R!$ unique priority orders among the classes and we need to determine the optimal fraction of time $\gamma_i^*$ the scheduler serves using the $i^{\t{th}}$ priority order. Thus, we need to solve a set of $R!+1$ nonlinear equations to determine $\gamma^*$. Hence, the computational complexity of determining the optimal scheduler is enormous, even at moderate number of classes, $R$. 

\subsubsection{Iterative optimization} \label{complexIter} 
\begin{figure}
\centering
\includegraphics[scale=0.45]{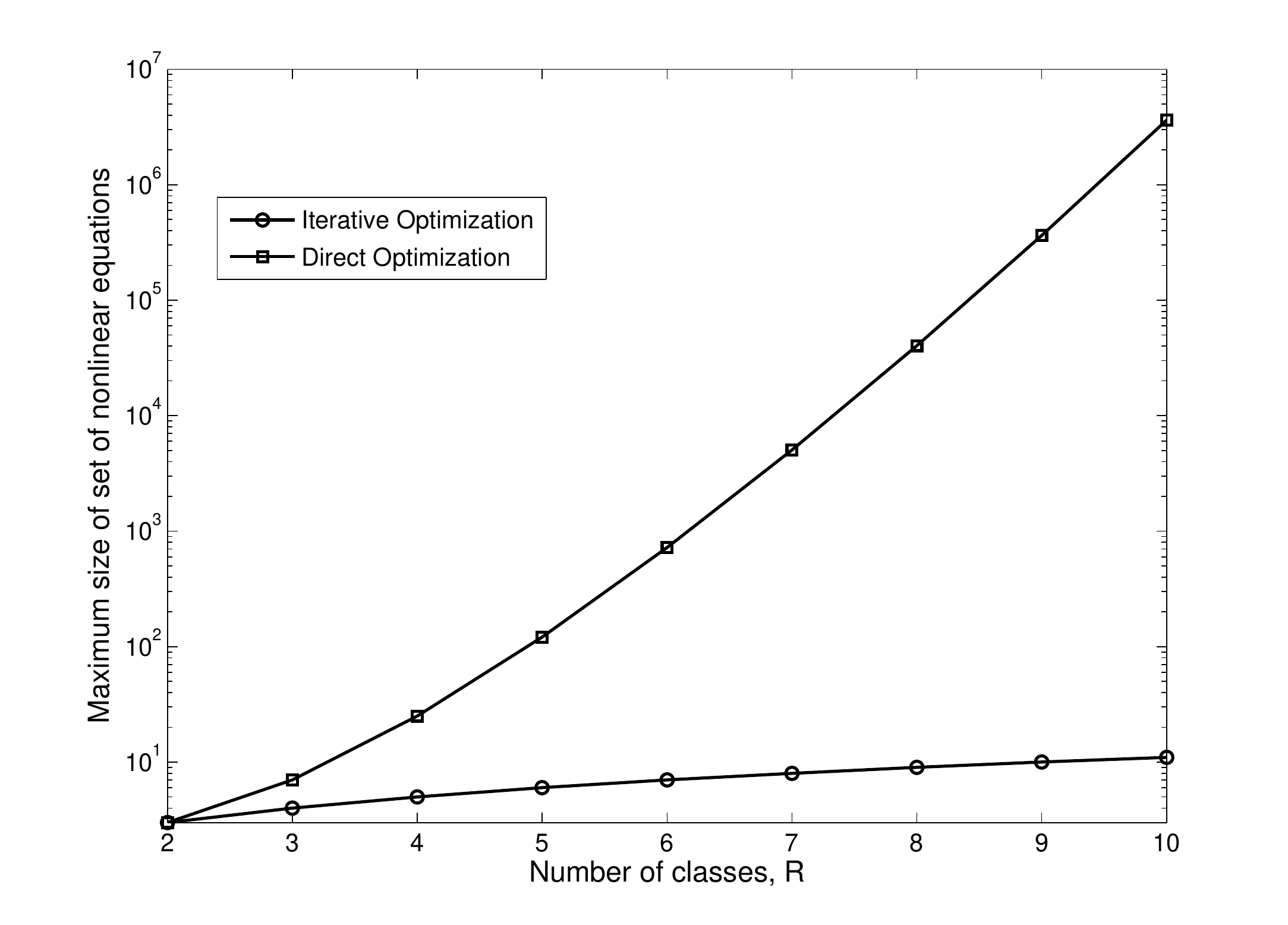}%
\caption{Complexity of iterative and direct optimization schemes.}
\label{complexityComparison} 
\end{figure}

In this case, we solve optimization problems for $r$ class system, $\forall~\mathcal{A}_r, \forall~ 2\leq r\leq R$. Since the priority order among the classes in $\mathcal{A}_r^\mathsf{c}$ does not effect the latency of classes in $\mathcal{A}_r$, the number of $r$ class optimization problems are $\binom{R}{r}$, each requiring us to solve a set of $r+1$ nonlinear equations. Thus, the iterative algorithm scales well with the number of classes by solving multiple but quite small optimization problems. The maximum number of simultaneous nonlinear equations is $R+1$ for $r=R$. This complexity reduction using iterative optimization is illustrated in Fig.~\ref{complexityComparison} that plots the maximum number of simultaneous nonlinear equations that needs to be solved in both schemes as the number of classes $R$ are increased. 

Clearly, the maximum complexity of an optimization problem for iterative optimization is negligible compared to the direct optimization, even though it requires us to solve a large number of optimization problems.

\section{Jointly Optimal MA-AS Scheduler}
\label{jointOptimal}
We now extend the proposed scheduler to design a jointly-optimal packet scheduler at $M$ MAs and AS. In our system model, we assumed a Poisson arrival process and constant packet size for each QoS class at each MA. Therefore, the queuing process for any class at any given MA is a M/D/1 process and hence the aggregated traffic for each class at AS becomes a G/D/1 process. 

\subsection{Centralized Optimization at M2M AS}
\label{centralOptimM2M}
We extend the analysis in Section~\ref{directOptimProb} to formulate a joint utility-maximization problem to be solved centrally at AS in a single stage. We first define the set $\mathcal{M} = \{1, 2, \cdots, M\}$. Let the $k^{\t{th}}$ MA adopt the $j^{\t{th}}$ non-preemptive priority order\footnote{We assume non-preemptive priority scheduling at each MA to prevent discontinuities in transmission of a packet over the wireless channel.} for $\gamma_{j,k}$  fraction of the time and result in average latency of $l_{i,j,k}^s$ for class $i$, $\forall~i\in \mathcal{R}, \forall~j \in \mathsf{R}$ and $\forall~k \in \mathcal{M}$. Then the average latency $l_i$ of class $i$ is given by,
\begin{equation}
\label{latEqSet_Centralized}
l_i = \sum_{j=1}^{R!} \gamma_{j} l_{i,j}^s + \sum_{k=1}^{M} w_{i,k} \sum_{j=1}^{R!} \gamma_{j,k} l_{i,j,k}^s ~\forall~i,
\end{equation}
where the first component represents the latency experienced at AS as in \eqref{latEqSet_direct} and the weight $w_{i,k} = \lambda_{ik}/\lambda_{i}$ represents the contribution of $k^{\t{th}}$ MA towards overall latency of class $i$. Let $\mathcal{H}_{i,j}^k$ denote the set of classes with higher priority than class $i$ in the $j^{\t{th}}$ priority order at $k^{\t{th}}$ MA. Using the existing result on latency of M/D/1 nonpreemptive priority system \cite{gallager}, we have,
\begin{align}
l_{i,j,k}^s &= \frac{1}{\mu_{ik}} + \frac{1}{\rho_{\mathcal{H}_{i,j}^k} (\rho_{\mathcal{H}_{i,j}^k}-\rho_{ik})} \sum_{m \in \mathcal{R}} \frac{\rho_{mk}}{2\mu_{mk}} ~ \forall~i, \label{lat_i_central}
\end{align}
where $\rho_{ik}=\lambda_{ik}/\mu_{ik}$ is the utilization factor for class $i$ at $k^{\t{th}}$ MA and $\rho_{\mathcal{H}_{i,j}}^k = 1-\sum_{m \in \mathcal{H}_{i,j}^k} \rho_{mk}$. 

A closed-form expression for $l_{i,j}^s$ exists only if we have a M/G/1 queuing system at AS and is given in \eqref{lat_i_direct}. However, due to the assumption of constant packet sizes, the queuing process at MAs and AS is M/D/1 and G/D/1 respectively. In this case, the latency $l_{i,j}^s$ can be determined by simulating the queuing processes.

Let $\hat{\gamma} = \{\gamma_j: j\in \mathsf{R}\}$ and $\hat{\gamma_k} = \{\gamma_{j,k}: j\in\mathsf{R}\}$. Then the joint scheduling policy is characterized by $\gamma = \{\hat{\gamma}, \hat{\gamma_{k}}: k \in \mathcal{M}\}$. Then, similar to Section~\ref{directOptimProb}, we obtain the following Centralized Optimization problem,
\begin{equation}
\begin{aligned}
&\operatorname*{max}_{\gamma} & & \text{log}\left(V(\gamma)\right) = \sum_{i=1}^R {\beta_i}\text{log}\left(U_i(\gamma)\right)  \\
&\text{s.t.} & &\gamma_j, \gamma_{j,k} \geq 0 ~\forall~j \in \mathsf{R}, \forall~k \in \mathcal{M}  \\
& & & \sum_{j=1}^{R!} \gamma_j = 1, \sum_{j=1}^{R!} \gamma_{j,k} = 1 ~\forall~j, ~\forall~k \in \mathcal{M}.
\end{aligned}
\label{centralOptim}
\end{equation}
Since this optimization problem has $(M+1)R!$ variables, it is prohibitively complex to solve, even at moderate values of $M$ and $R$. 
\subsection{Distributed Optimization}
\label{DistributedOptim}
Since the complexity of the centralized optimization is prohibitively high, we subdivide it into $M+1$ optimization problems to be solved separately at AS and the $M$ MAs in an iterative fashion. At the start of each iteration, each MA receives information about the locally optimal scheduler\footnote{Here locally optimal scheduler means the optimal scheduler at a MA or AS determined by fixing the scheduling decisions at other MAs and AS.} at the other MAs and AS determined from previous iteration. Then each MA simultaneously solves the centralized optimization problem in \eqref{centralOptim} by fixing the scheduling decisions for other MAs and AS and solving for its locally optimal scheduler using the iterative optimization framework presented in Section~\ref{iterativeOptim}. After this the AS receives information about the locally optimal scheduler from all the MAs to determine its locally optimal scheduler. We can completely decouple the optimization problems at MAs and AS, given that the arrival process at AS is represented by a Poisson process. 

We now describe the optimization sub-problem at $m^{\t{th}}$ MA and AS at ${n}^{\t{th}}$ iteration.  
\subsubsection{Optimization at $m^{\t{th}}$ MA} \label{optMA} We now refer to the iterative optimization framework presented in Section~\ref{iterativeOptim}. For a given set $\mathcal{A}_r$ of $r$ classes, if the $m^{\t{th}}$ MA serves class $i \in \mathcal{A}_r$ with highest priority for $\alpha_{i}$ fraction of time, then using \eqref{latEqSet} and \eqref{latEqSet_Centralized}, the average latency $l_i$ of class $i$ is,
\begin{align}
l_i &= \underbrace{\sum_{j=1}^{R!} \gamma_{j}^{n-1} l_{i,j}^s + \sum_{k=1, k\neq m}^{M} w_{i,k} \sum_{j=1}^{R!} \gamma_{j,k}^{n-1} l_{i,j,k}^s}_{\psi_{im}^{n-1}}  \nonumber \\ &+ w_{i,m}(\alpha_{i,m} l_{i,i,m} + \sum_{j \in \mathcal{A}_r, j\neq i} \alpha_{j,m} l_{i,j,m}^*) ~\forall~i \in \mathcal{R}, \nonumber\\ 
&= \psi_{im}^{n-1} + w_{i,m}(\alpha_{i,m} l_{i,i,m} + \sum_{j \in \mathcal{A}_r, j\neq i} \alpha_{j} l_{i,j,m}^*), \label{latEqSet_DistMAk}
\end{align}
where $\gamma_{j}^{n-1}$ and $\gamma_{j,k}^{n-1}$ denotes the optimal value of $\gamma_{j}$ and $\gamma_{j,k}$ at ${(n-1)}^{\t{th}}$ iteration. Here $\alpha_{i,m},~l_{i,i,m}$ and $l_{i,j,m}^*$ have same meaning as $\alpha_i, l_{i,i}$ and $l_{i,j}^*$ in Section~\ref{iterativeOptim}, but are defined for $m^{\t{th}}$ MA. Similarly, we define $\alpha_{\mathcal{A}_r}^m = \{\alpha_{i,m}: i\in \mathcal{A}_r\}$.

Since we assume the packet size is constant, the queuing process at AS is G/D/1. Therefore, the arrival process for each class and hence the latency terms $l_{i,j}^s$ at AS are functions of the scheduling policy $\hat{\gamma}_k$ at the $k^{\t{th}}$ MA $\forall~k$. Given that the arrival process for each class at AS is a Poisson process, we can decouple the optimization problems at each MA and AS as $l_{i,j}^s$ is independent of the scheduling policy $\hat{\gamma}_k$ and thus the term $\psi_{im}^{n-1}$ can be treated as a constant. 

We note that \eqref{latEqSet_DistMAk} is similar to \eqref{latEqSet} except for the additive constant $\psi_{im}^{n-1}$ and multiplicative constant $w_{i,m}$. Therefore using the procedure in in Section~\ref{iterativeOptim}, the resultant $r$-class optimization problem at $m^{\t{th}}$ MA can also be easily shown to be convex from Theorem~\ref{Theorem1} and thus solved iteratively using Algorithm~\ref{propScheduler}. Finally, the optimal solution at each iteration of Algorithm~\ref{propScheduler} is aggregated to determine the fraction of time the scheduler at $m^{\t{th}}$ MA uses each of the $R!$ priority orders during the ${n}^{\t{th}}$ iteration, i.e., $\hat{\gamma}_m^n =\{\gamma_{j,m}^{n}: j \in \mathsf{R}\}$.
 

\subsubsection{Optimization at AS}
 For a given set $\mathcal{A}_r$ of $r$ classes, if the AS serves class $i \in \mathcal{A}_r$ with highest priority for $\alpha_{i}$ fraction of time, then using \eqref{latEqSet} and \eqref{latEqSet_Centralized}, the average latency $l_i$ of class $i$ at ${n}^{\t{th}}$ iteration is,
\begin{equation}
\label{latEqSet_DistAS}
\begin{aligned}
l_i &= \alpha_{i} l_{i,i} + \sum_{j \in \mathcal{A}_r, j\neq i} \alpha_{j} l_{i,j}^* + \underbrace{\sum_{k=1}^{M} w_{i,k} \sum_{j=1}^{R!} \gamma_{j,k}^{n} l_{i,j,k}^s}_{\psi_{i}^{n}},\\
&= \alpha_{i} l_{i,i} + \sum_{j \in \mathcal{A}_r, j\neq i} \alpha_{j} l_{i,j}^* + \psi_{i}^{n} ~\forall~i \in \mathcal{R},
\end{aligned}
\end{equation}
where the last term $\psi_{i}^{n}$ is a constant. Again, for reasons mentioned earlier, the resultant $r$-class optimization problem at AS is convex and is solved iteratively using Algorithm~\ref{propScheduler}. Finally, the optimal solution at each iteration of Algorithm~\ref{propScheduler} is aggregated to determine the fraction of time the scheduler at AS uses each of the $R!$ priority orders during the ${n}^{\t{th}}$ iteration, i.e., $\hat{\gamma}^n =\{\gamma_{j}^{n}: j \in \mathsf{R}\}$.

\begin{figure}
\centering
\includegraphics[scale=0.55]{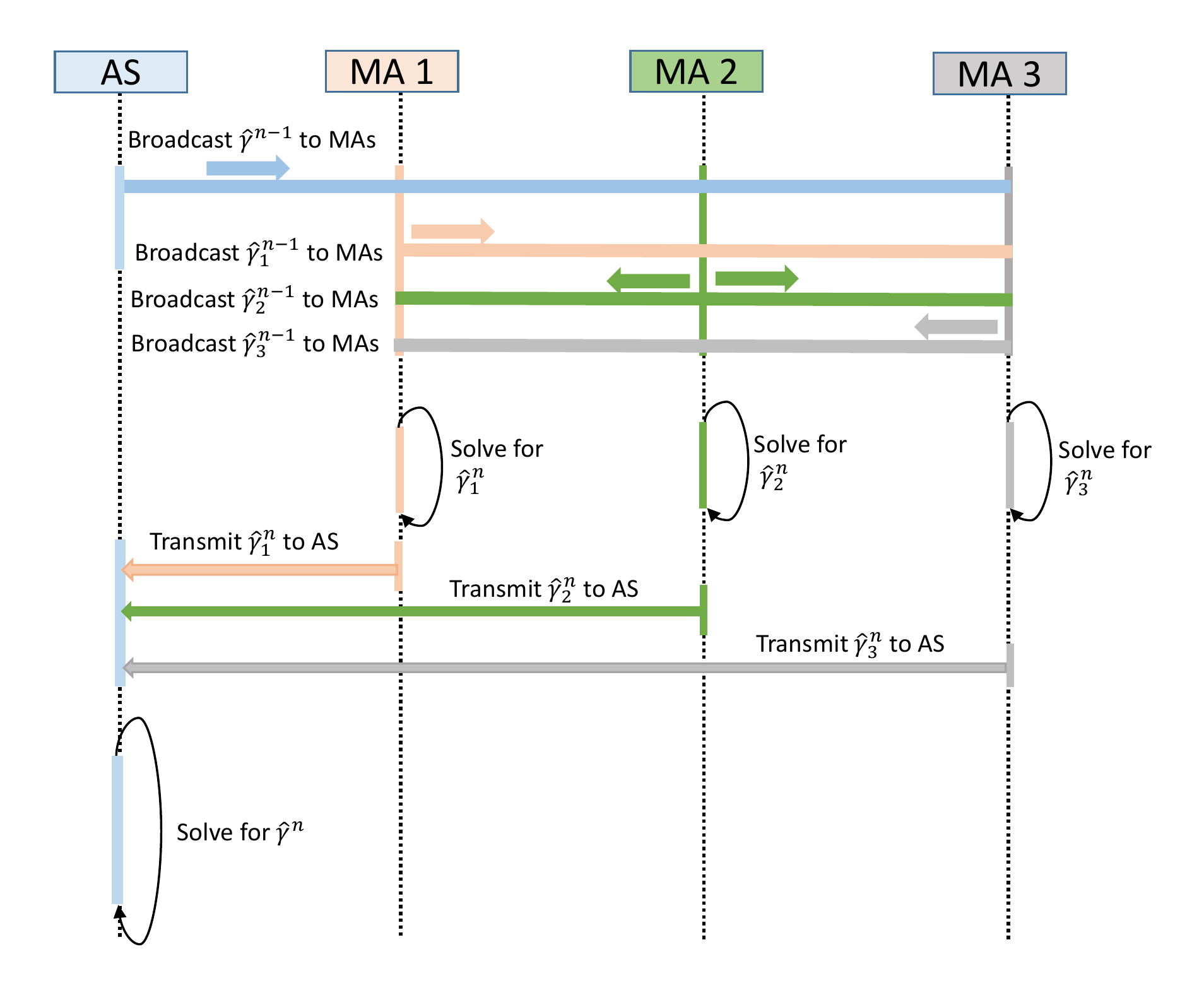}%
\caption{Sequence diagram illustrating the Distributed optimization process at $n^{\t{th}}$ iteration.}
\label{seqDiagram_DO}
\end{figure}

The distributed optimization process at $n^{\t{th}}$ iteration for $M=3$ is illustrated in Fig.~\ref{seqDiagram_DO}. A simple way to initialize the local scheduling decisions, prior to the $1^{\t{st}}$ iteration, is by setting $\gamma_j^0 = 1/R!~\forall~j \in \mathsf{R}$ for AS and $\gamma_{j,k}^0 = 1/R!~\forall~j\in\mathsf{R}$ for the $k^{\t{th}}$ MA $\forall~k\in\mathcal{M}$.

\subsection{Complexity Analysis}
We now compare the complexity of the centralized and distributed optimization problem for a M2M uplink with $M$ MAs and $R$ classes.\\
\subsubsection{Centralized Optimization} Here, we formulate a single-stage optimization problem at AS with $(M+1)R!$ variables as can be noted from Eq.~\eqref{centralOptim}. On solving the Lagrangian dual problem similar to the procedure in Section~\ref{optSch}, the problems reduces to solving a set of $(M+1)R!+M+1$ nonlinear equations simultaneously.\\
\subsubsection{Distributed Optimization}\label{cplx_distOpt}
In the distributed optimization framework, at each iteration, we adopt the iterative optimization approach to determine the locally optimal scheduler at the MAs and AS. Again as discussed in Section~\ref{complexIter}, we solve $\binom{R}{r}(M+1)$ optimization problems for $r$-class system ($\forall~ 2\leq r\leq R$), each requiring to solve a set of $r+1$ simultaneous equations. However, the maximum number of simultaneous nonlinear equations solved at any MA or AS is still $R+1$, which is much smaller than that needed for centralized optimization.

We now analyze the overhead due to information exchange among the MAs and AS during each iteration. We note that, the set $\alpha_{\mathcal{A}_r}$ and $\alpha_{\mathcal{A}_r}^k~\forall~\mathcal{A}_r,~\forall~r$ is a complete characterization of $\hat{\gamma}$ and $\hat{\gamma}^k$ respectively. So we can either choose to broadcast $\alpha_{\mathcal{A}_r}, \alpha_{\mathcal{A}_r}^k$ or $\hat{\gamma}, \hat{\gamma}^k$. 

\emph{Case 1: Broadcast $\alpha_{\mathcal{A}_r}, \alpha_{\mathcal{A}_r}^k$}\\
Each node solves $\binom{R}{r}$ $r$-class optimization subproblems. The total number of optimization variables for $R$-class problem at any node is, 
\begin{align}
N_R &= \sum_{r=2}^R \binom{R}{r}(r-1), \label{N_R_exp}\\
    &= (R-1)(2^{R-1}-1)-2^{R-1}+R,
\end{align}
where in Eq.~\ref{N_R_exp}, we consider $r-1$ variables for $r$-class problem as the sum of the optimization variables in each subproblem is constant (equal to $1$). Therefore, the total number of variables transmitted over-the-air due to $M$ MAs and AS is $N_1 = (M+1)N_R$.

\emph{Case 2: Broadcast $\hat{\gamma}, \hat{\gamma}^k$} \\
In this case, each node broadcasts $R!-1$ variables. Thus a total of $N_2 = (M+1)(R!-1)$ variables are transmitted.

\begin{figure}
\centering
\includegraphics[scale=0.45]{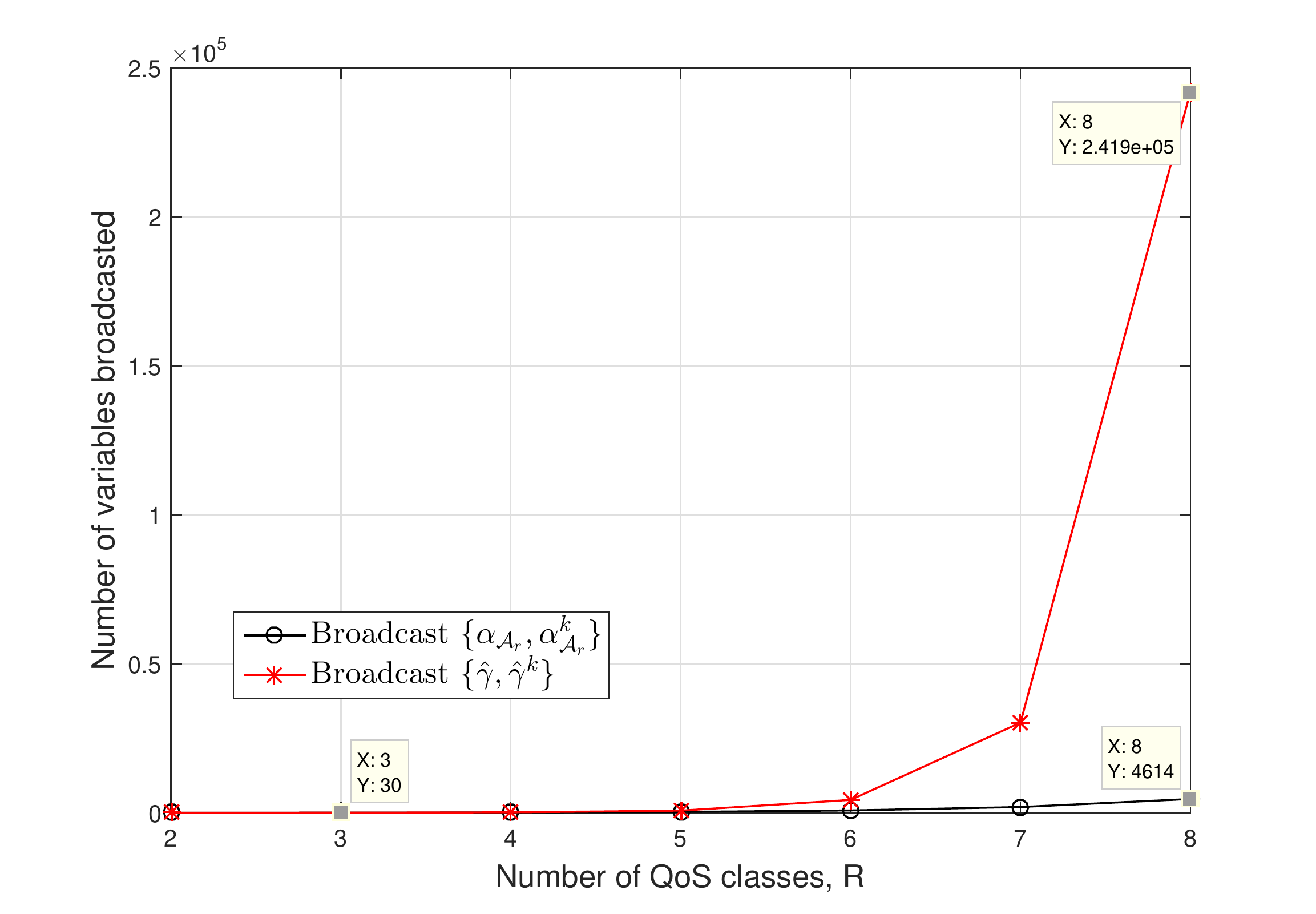}%
\caption{Information exchange overhead as $R$ is varied.}
\label{plotNumMessage} 
\end{figure}

Fig.~\ref{plotNumMessage} plots the number of optimization variables broadcasted in each case as $R$ is varied. We note that the number of variables in two cases are same upto $R=3$ and then it grows drastically in case $2$. At $R=8$, the number of variables in case $2$ is roughly $52$ times more than that in case $1$. Therefore, we choose to broadcast $\alpha_{\mathcal{A}_r}, \alpha_{\mathcal{A}_r}^k$ for information exchange in each iteration.

\vspace{10pt}

\section{Joint Subcarrier Allocation and Packet Scheduler}
\label{jointSubcarrierScheduler}
We now extend our work to jointly determine the delay-optimal subcarrier allocation and packet scheduling for the MA-AS system.

\subsection{Centralized Optimization at M2M AS}
\label{central_jointChanScheduler}
We now formulate the joint optimization problem to be solved centrally at the M2M Application Server (AS). The problem formulation follows on similar line as in Section~\ref{centralOptimM2M}. The difference is that now the service rates at MAs are not fixed but depends on the number of subcarriers ($N_k$) assigned to the MAs which is determined by solving the following joint optimization problem.

\begin{equation}
\begin{aligned}
\operatorname*{max}_{\gamma_j,\gamma_{j,k},N_k} &   \sum_{i \in \mathcal{R}} {\beta_i}\text{log}\left(U_i(\gamma_j,\gamma_{j,k},N_k)\right)  \\
\text{s.t.} & \gamma_j, \gamma_{j,k} \geq 0 ~\forall~j \in \mathsf{R}, \forall~k \in \mathcal{M}  \\
 & N_k \in \mathbb{Z}^{+}, \forall~k \in \mathcal{M}  \\
 & \sum_{j \in \mathsf{R}} \gamma_j = 1, \sum_{k \in \mathcal{M}} N_k = N \\
 & \sum_{j \in \mathsf{R}} \gamma_{j,k} = 1 ~\forall~k \in \mathcal{M}.
\end{aligned}
\label{centralOptim_ChanSch}
\end{equation}
 
We note that $\gamma_j, \gamma_{j,k}$ are continuous variables whereas $N_k$ are integer variables. Therefore, this optimization problem is a mixed integer nonlinear programming (MINLP) problem and are in general shown to be NP-hard. Furthermore, this optimization problem has a total of $(M+1)R!+N$ variables, which is huge even at moderate values of $M$ and $R$. Thus, the centralized joint optimization problem is prohibitively complex to solve and hence necessitates the need of alternative low-complexity optimization procedures.

Also, as pointed out earlier in Section~\ref{centralOptimM2M}, there is no closed form expression for average latency of classes at AS if packet size is assumed constant, i.e.,for a G/D/1 queuing process at AS. Resorting to numerical solution for approximating the latency expression becomes extremely difficult here due to variable MA service rate. Therefore, we assume the packet size for each class to be geometrically distributed, resulting in an exponential service time distribution and thus M/M/1 arrival process at MAs and AS. We show later in Fig.~\ref{jointSysUtil} that the approximation of exponential service distribution results in a near-optimal delay performance.

We next present a novel low-complexity iterative optimization framework that is implemented in a distributed fashion across the MAs and AS. 

\subsection{Proposed Distributed Optimization Framework}
\label{distrib_ChannelScheduler}
We solve the joint channel assignment and packet scheduling problem into two sequential stages. Stage $1$ involves solving a distributed optimization sub-problem at each MA to determine the optimal subcarrier assignment to MAs. This is then fed as input to Stage $2$ which determines the jointly optimal scheduling problem at MAs and AS while fixing the subcarrier assignment from Stage $1$. Since the Stage $2$ is same as discussed previously in Section~\ref{DistributedOptim}, here we focus only on Stage $1$.

\subsubsection{Stage 1: Optimal Subcarrier Allocation to MAs}
We initialize Stage $1$ with a \emph{minimally-feasible} assignment of subcarrier which is defined as follows,
\begin{Def}\label{feasibleDef}
Consider a M2M uplink with $M$ MAs, $R$ traffic classes with $\lambda_{ik}$ and $\mu_{ik} = \frac{N_k C_s}{s_i}$ denoting the arrival and service rate of $i^{\t{th}}$ class at $k^{\t{th}}$ MA respectively. Let $\mathcal{N}=\{N_1, N_2, \cdots, N_m\}$ denote the set of any unconstrained integer subcarrier assignment. Then the minimally-feasible subcarrier assignment $\mathcal{N}^{\t{mf}}=\{N_1^{\t{mf}}, N_2^{\t{mf}}, \cdots, N_m^{\t{mf}}\}$ is defined such that if there exists a subcarrier assignment $\mathcal{N}$ with $N_k<N_k^{\t{mf}}$ for any $k \in \mathcal{M}$, then $\sum_{i\in \mathcal{R}} \frac{\lambda_{ik}}{\mu_{ik}}>1$ and thus results in a unstable queuing system at $k^{\t{th}}$ MA. 
\end{Def} 
Thus the \emph{minimally-feasible} assignment is the smallest possible subcarrier assignment at each MA beyond which it would result in an unbounded queue at that MA. It is straightforward to prove that the \emph{minimally-feasible} assignment is unique.

Next, for the \emph{minimally-feasible} assignment, the $k^{\t{th}}$ MA $\forall~k$ solves a locally optimal packet scheduling problem (\textbf{DO-k}) without any knowledge of scheduling policy at other MA or AS. We enable this be setting $l_i$ as the average latency of $i^{\t{th}}$ class at $k^{\t{th}}$ MA as,
\begin{equation}
l_i =  w_{i,k} \sum_{j \in \mathsf{R}} \gamma_{j,k} l_{i,j,k} ~\forall~i \in \mathcal{R},
\end{equation}
Therefore, we obtain the following optimization problem at $k^{\t{th}}$ MA.
\begin{equation}
\begin{aligned}
\textbf{DO-k:}&\operatorname*{max}_{\gamma_{j,k}} & & Z_k^{\t{mf}} = \sum_{i \in \mathcal{R}} {\beta_i}\text{log}\left(U_i(\gamma_{j,k},N_k^{\t{mf}})\right)  \\
&\text{s.t.} & & \gamma_{j,k} \geq 0 ~\forall~j \in \mathsf{R}  \\
& & & \sum_{j \in \mathsf{R}} \gamma_{j,k} = 1.
\end{aligned}
\label{mfaOptimK}
\end{equation}
Let ${\hat{Z}}_k^{\t{mf}}$ denote the value of objective for $k^{\t{th}}$ MA at the optimal solution and let ${\hat{Z}}^{\t{mf}} = \{{\hat{Z}}_1^{\t{mf}}, {\hat{Z}}_2^{\t{mf}}, \cdots, {\hat{Z}}_M^{\t{mf}}\}$. To further reduce the computation complexity associated with the optimization problem $\textbf{DO-k}$, we adopt the iterative optimization scheme discussed in our previous work.

Next, we assign the remaining subcarriers one by one to the MA that results in the maximum increase in system utility\footnote{ Since the capacity of each subcarrier is same for AWGN case, it is irrelevant which subcarrier gets assigned to a MA; rather what matters is the number of subcarriers assigned to each MA.} Thus, the subcarrier assignment is an iterative process with $\tilde{N} = N-\sum_{k \in \mathcal{M}} N_k^{\t{mf}}$ iterations. During each iteration, we solve the optimization problem at $k^{\t{th}}$ MA  similar to Eq.~\eqref{mfaOptimK} (again using the iterative approach in our previous work) only if it is assigned a subcarrier in the previous iteration, thus avoiding solving unnecessary optimization problems. The proposed subcarrier assignment algorithm is described in Algorithm~\ref{propChannelAss}. The inputs to the algorithm are the \emph{minimally-feasible} subcarrier assignment $\mathcal{N}^{\t{mf}}$ and the resultant optimal objective value at MAs ${\hat{Z}}^{\t{mf}}$. The output is the number of subcarriers assigned to each MA, $\mathcal{N}$.

\begin{algorithm}[!h]
\caption{Proposed Subcarrier Assignment Algorithm}
\label{propChannelAss}
\small{
\begin{algorithmic} 
	\Function{optChAss}{$\mathcal{N}^{\t{mf}}, {\hat{Z}}^{\t{mf}}$}
	     \State Set $\mathcal{N} = \mathcal{N}^{\t{mf}}$, ${\hat{Z}} = {\hat{Z}}^{\t{mf}}$.
	     \State Set $\tilde{N} = N-\sum_{k \in \mathcal{M}} N_k^{\t{mf}}$.
	      \State Set update flag $f_k = 1~ \forall~ k \in \mathcal{M}$.
		 \For{$i = N-\tilde{N}+1: N$} \Comment{Iterate over remaining subcarriers}		                 	        \For{$k=1:M$} \Comment{Iterate over each MA}
			  \If{$f_k = 1$}
			    \State Solve $\textbf{DO-k}$ with $N_k+1$ subcarriers.
			    \State Calculate the differential utility $\Delta {\hat{Z}}_k$.  
			  \EndIf					
			\EndFor	
			\State Find the MA, $k^{*}$, with largest $\Delta {\hat{Z}}_k$.	
			\State Set $f_k = 0, \forall~ k \neq k^{*}$.
			\State Set $N_{k^{*}}=1~ \&~ {\hat{Z}}_{k^{*}} = {\hat{Z}}_{k^*}+\Delta {\hat{Z}}_{k^{*}}$.
		\EndFor
		
		\Return $\mathcal{N}$.
		\EndFunction
	\end{algorithmic}	
	}
\end{algorithm}

\subsubsection{Stage 2: Optimal Packet Scheduler at MAs and AS}
The Stage $2$ problem involves determining the jointly optimal packet scheduling scheme at MAs and AS while fixing the subcarrier allocation as the output of Stage $1$. This problem can be solved with low-complexity in a distributed and iterative fashion across MAs and AS as discussed in Section~\ref{DistributedOptim}.

\subsection{Complexity Analysis}
We now compare the complexity of the centralized and distributed frameworks for joint subcarrier assignment and packet scheduler in a M2M uplink with $M$ MAs, $R$ classes and $N$ subcarriers.

\subsubsection{Centralized Optimization} Here, we formulate and solve a single-stage MINLP optimization problem at AS with $(M+1)R!+N$ variables. Since this is a NP-hard problem, the computational complexity is prohibitively high even at moderate values of $M$, $R$ and $N$.

\subsubsection{Distributed Optimization}
Here we discuss the computational complexity of proposed subcarrier assignment algorithm (Stage $1$) detailed in Algorithm~\ref{propChannelAss}. The complexity of Stage $2$ is same as described in Section~\ref{cplx_distOpt}. Algorithm~\ref{propChannelAss} is initialized with a \emph{minimally-feasible} subcarrier assignment $\mathcal{N}^{\t{mf}}$ as defined in Definition~\ref{feasibleDef}. 
The second input argument ${\hat{Z}}^{\t{mf}}$ is the result of solving the local optimization problem \textbf{DO-k} (defined in Eq.~\ref{mfaOptimK}) at the $k^{\text{th}}$ MA $\forall k\in\mathcal{M}$ with $N_k^{mf}$ subcarriers. Then, we allocate the remaining $\tilde{N}$ subcarriers one-by-one to the MAs. The first iteration requires $k^{\text{th}}$ MA $\forall k\in\mathcal{M}$ to solve the optimization problem \textbf{DO-k} with $N_k^{mf}+1$ subcarriers. Thus the complexity of first iteration is same as that of determining ${\hat{Z}}^{\t{mf}}$. Then in each of $\tilde{N}-1$ iterations, we solve \textbf{DO-k} only at the MA which is allocated an new subcarrier in the previous iteration. 

Putting everything together, we solve $2M+\tilde{N}-1$ instances of the optimization problem \textbf{DO-k} using the iterative procedure described in Section~\ref{iterativeOptim}. The computational complexity of each problem is described previously in Section~\ref{complexIter} and requires solving a maximum of $R+1$ simultaneous nonlinear equations.

We next analyze the information exchange overhead in Stage $1$ of distributed optimization framework. The corresponding analysis for Stage $2$ has been done in Section~\ref{cplx_distOpt}. Based on the channel conditions and arrival rate of different classes at each MA, the AS determines the \emph{minimally-feasible} subcarrier assignment and feeds back this information to the MAs. Then the procedure for allocating the remaining $\tilde{N}$ subcarriers is done in $\tilde{N}$ iterations. In the first iteration, all MAs compute and transmit their corresponding $\Delta {\hat{Z}}_k$ to AS. The AS then determines the MA index with largest $\Delta {\hat{Z}}_k$ and broadcasts this MA index to the MAs. On receiving the message from AS, the MA with the allocated subcarrier updates its $\Delta {\hat{Z}}_k$ for next iteration, while other MAs simply ignore the AS message. Thus for each of the remaining $\tilde{N}-1$ iterations, the $k^{\text{th}}$ MA updates and transmits $\Delta {\hat{Z}}_k$ to AS only if it was assigned a subcarrier in the last iteration. 

Putting everything together, we note that the AS transmits $M$ messages with initial subcarrier assignment to each MA and $\tilde{N}$ messages with MA index that is allocated next subcarrier. The The total number of $\Delta {\hat{Z}}_k$ values transmitted by all MAs to the AS are $M+\tilde{N}-1$ .

\section{Numerical Results}
\label{Results_M2M2}
We now use Monte-Carlo simulations to compare the system utility and delay jitter performance of proposed scheduler against various state-of-the-art schedulers, namely WRR, WFS, max-weight scheduler and priority scheduling schemes. We consider $4$ QoS classes classes with arrival rate as $[\lambda_2, \lambda_3, \lambda_4] = [0.02, 0.04, 0.05]~\text{s}^{-1}$ , packet size as $[s_1, s_2, s_3, s_4] = [143, 111, 83, 67]$~B and service rate $r=100$~B/s. The utility function parameters are as follows: $[a_1, a_2, a_3, a_4] = [0.35, 0.4, 0.45, 0.7]~\text{s}^{-1}$, $[b_1, b_2, b_3, b_4] = [2, 1.9, 1.8, 1]$~s and $[\beta_1, \beta_2, \beta_3, \beta_4] = [0.35, 0.45, 0.5, 0.8]$. For WRR, the weights for class $i$ are inversely proportional to the delay requirement (set as $b_i+\frac{4}{a_i}$) and the packet size $s_i$. After normalizing to integer values, we get the weights for WRR as $[52, 76, 112, 223]$. For WFS, the weights are calculated similarly, except that we do not factor in the packet sizes. The weights for WFS are $[745, 840, 936, 1489]$.

\subsection{Multiclass Scheduler at AS}
\begin{figure}
\centering
\includegraphics[scale=0.5]{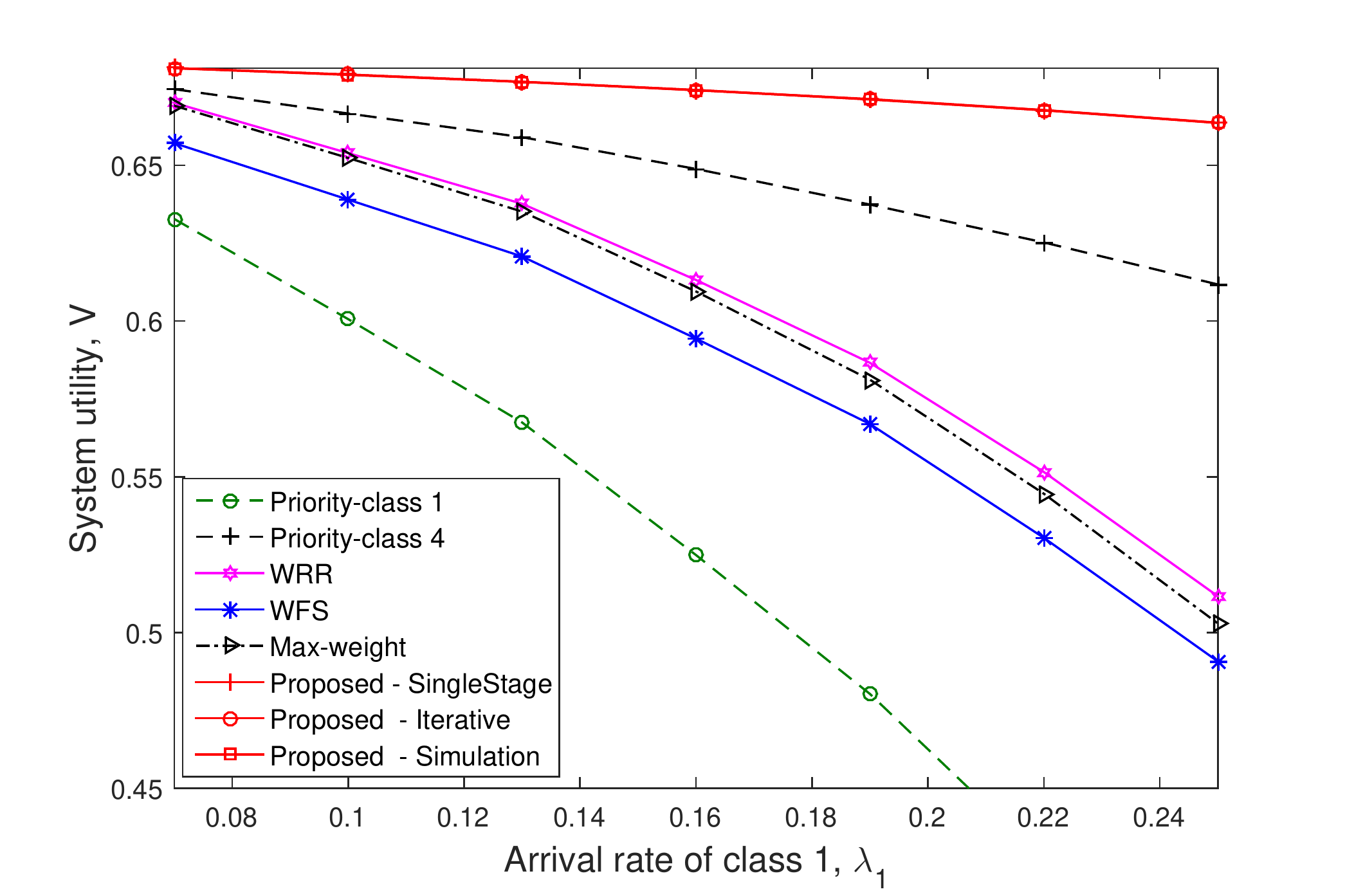}%
\caption{Comparison of system utility for $R=4$ classes.}
\label{sysUtility4class_AS} 
\end{figure}

Fig.~\ref{sysUtility4class_AS} compares the delay-performance of various scheduling schemes by plotting system utility metric as a function of $\lambda_1$. Firstly, we note that the theoretical result (both single-stage and iterative optimization) for proposed scheduler matches perfectly with the simulation, thus validating the correctness of our analysis. As expected, the system utility decreases with increase in $\lambda_1$ for all schedulers due to larger queuing delays. However, the proposed scheduler always outperforms other schedulers and the performance gap widens with increasing $\lambda_1$. This is because it prioritizes service to classes that are delay-sensitive and have greater impact of system utility. Therefore, the delay performance of proposed scheduler does not decrease significantly with increase in $\lambda_1$ due to the low-priority delay tolerant traffic of class $1$. For the same reason, we note that prioritizing class $1$ results in worst performance and priority to class $4$ performs really well and is second to only the proposed scheduler. This is because it gives equal priority to traffic of rest of the classes rather than prioritizing them based on their delay-sensitivity. 

\vspace{10pt}
The performance of WRR and WFS is bad despite assigning higher weight to delay-sensitive traffic due to the usage of non-optimal weights. Even if we had used optimal weights, the performance of WRR and WFS would still remain sub-optimal because of their goal to achieve weighted fairness among different classes. Lastly,  max-weight scheduling increasingly prioritizes class $1$ over other classes as $\lambda_1$ increases due to its largest average queue size. Since class $1$ is most delay-tolerant, this results in sharp performance loss with increasing $\lambda_1$.

\begin{figure}
\centering
\includegraphics[scale=0.45]{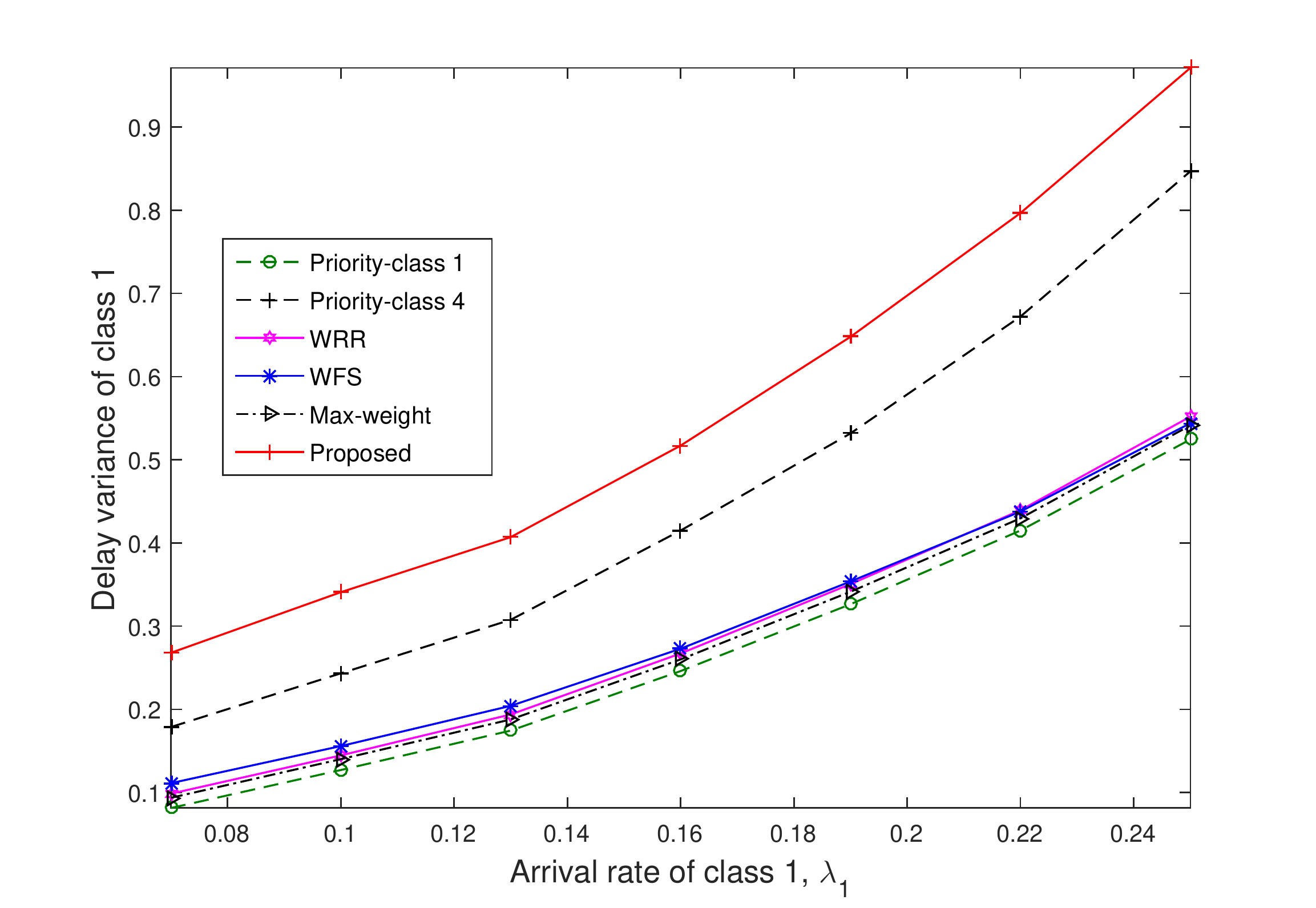}%
\caption{Delay variance of class 1 for different schedulers.}
\label{delayVarClass1_AS} 
\end{figure}

\begin{figure}
\centering
\includegraphics[scale=0.45]{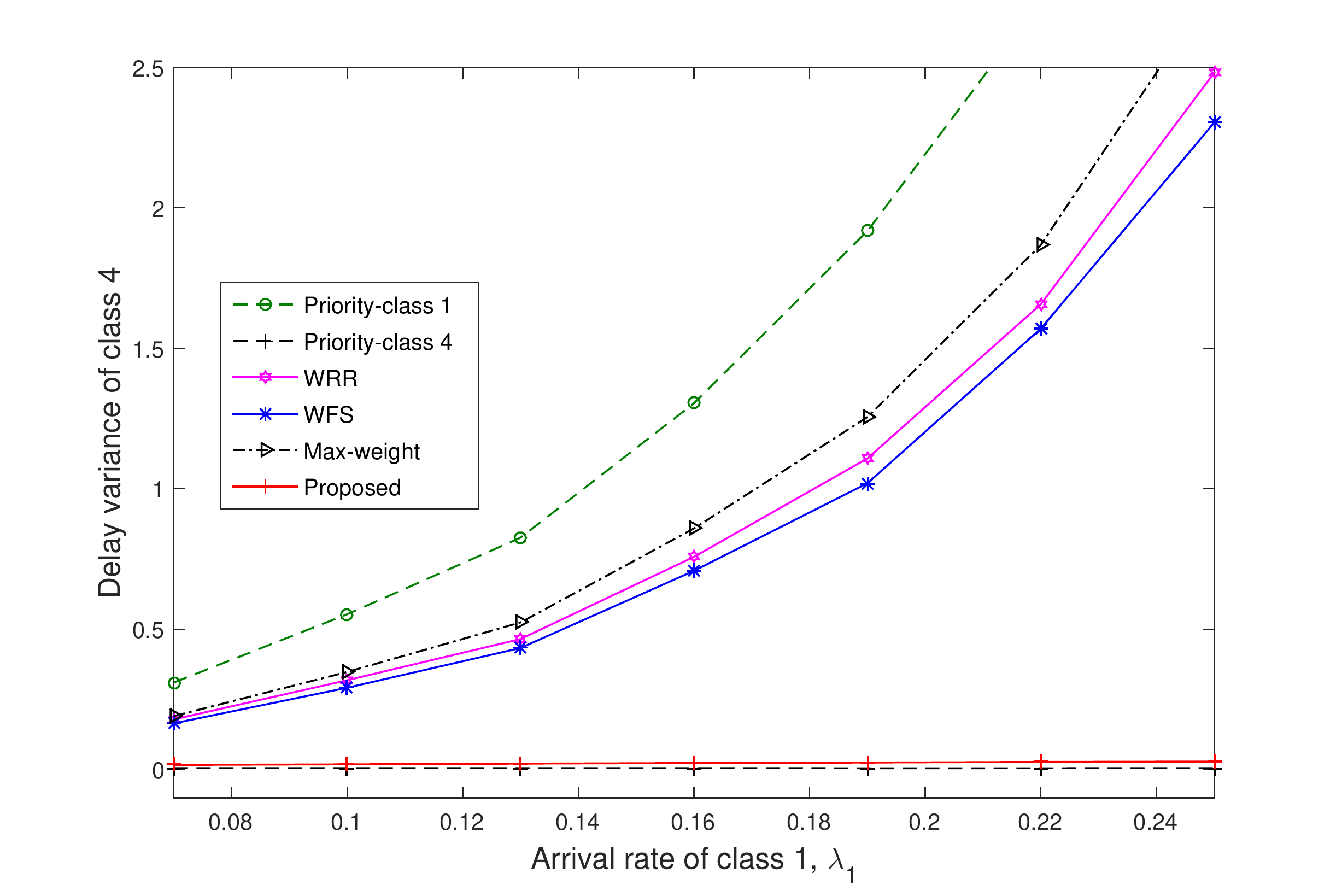}%
\caption{Delay variance of class 4 for different schedulers.}
\label{delayVarClass4_AS} 
\end{figure}

A low packet delay jitter, in general, is a desirable feature for scheduling policies. This is all the more important for delay-sensitive traffic where high delay jitter can lead to packets exceeding their delay budget. Therefore, we study the delay jitter (as measured by packet delay variance) of class $1$ and $4$ for different schedulers as shown in Fig.~\ref{delayVarClass1_AS} and ~\ref{delayVarClass4_AS} respectively. As suggested by intuition, we note that prioritizing traffic of a class results in its minimum delay variance. 

As desired, the proposed scheduler results in near-minimal delay variance for the delay-sensitive class $4$, far better than delay variance for other schedulers. This is because it gives high priority to class $4$. Although, the proposed scheduler performs poorly on delay variance of class $1$, this is not of grave concern due to its delay-tolerant and is also assigned least importance towards system utility. However, if low delay variance of class $1$ is important, it can still be achieved at the cost of higher average delay of class $1$, by using a delay-jitter regulator for class $1$ at AS. Lastly, we infer that the proposed scheduler and priority scheduler always result in the extreme delay variances for classes. On the contrary, WFS and WRR due to their round robin service operation, results in moderate delay-variance for all classes.

\subsection{Joint Multiclass Scheduler at MAs and AS}
Next we numerically evaluate the joint scheduling problem with $M=4$ MAs and $R=4$ classes. The utility parameters of each class, i.e., $a$, $b$ and $\beta$ remain unchanged. The arrival rate for different classes at each MA is expressed as a matrix given below,
\[ \left[ \begin{array}{cccc}
\lambda_{11} & \lambda_{12} & \lambda_{13} & \lambda_{14}\\
\lambda_{21} & \lambda_{22} & \lambda_{23} & \lambda_{24}\\
\lambda_{31} & \lambda_{32} & \lambda_{33} & \lambda_{34}\\
\lambda_{41} & \lambda_{42} & \lambda_{43} & \lambda_{44}
\end{array} \right] =\left[ \begin{array}{cccc}
5 & 4 & 4 & 3 \\
4 & 1 & 3 & 7 \\
3 & 8 & 5 & 2 \\
1 & 3 & 1 & 4
\end{array} \right] \times 10^{-2}~\text{s}^{-1}.\] 

The service rate at AS and MAs is $[r, r_1, r_2, r_3, r_4] =  [40, 34, 20, 60, 30]$~B/s and packet sizes for $4$ classes are $[s_1, s_2, s_3, s_4] = [14, 11, 8, 6]$~B.

\begin{figure}
\centering
\includegraphics[scale=0.5]{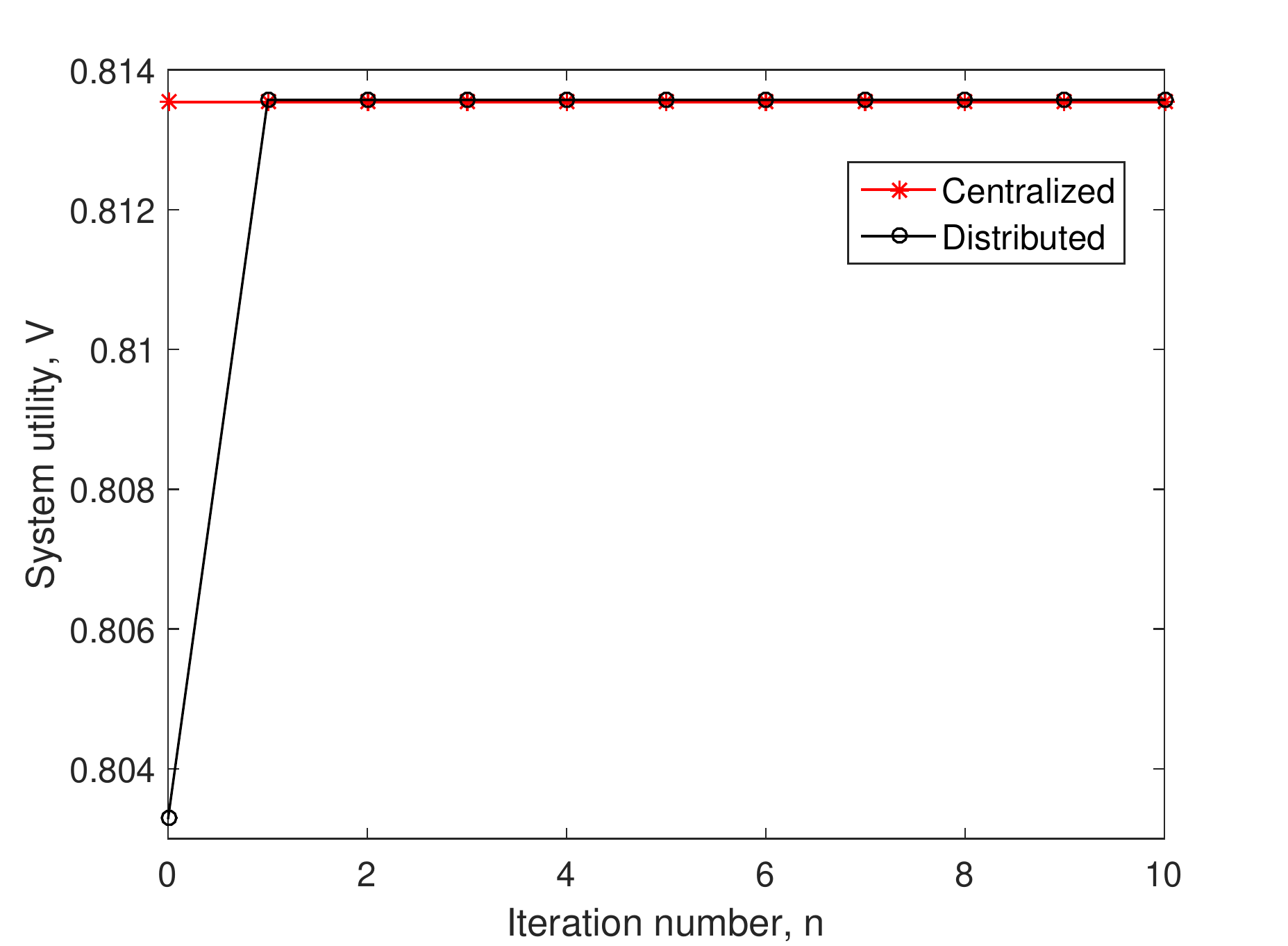}%
\caption{Convergence performance of distributed optimization for geometric packet size distribution.}
\label{convergenceDistCent} 
\end{figure}

Now, we expect the distributed optimization to be optimal when the arrival process at AS is actually Poisson. Using Burke's reversibility theorem \cite{Burke56}, this implies that the queuing process at each AS should be M/M/1, which in turn translates to geometric packet size distribution for each class. Using Fig.~\ref{convergenceDistCent}, we verify that this is indeed the case. We note that the system utility for distributed optimization converges quickly (in just a single iteration) to the centralized optimization result. Here, we had initialized $\gamma_j^0 = 1/R!~\forall~j \in \mathsf{R}$ for AS and $\gamma_{j,k}^0 = 1/R!~\forall~j\in\mathsf{R}$ for the $k^{\t{th}}$ MA $\forall~k\in\mathcal{M}$, assuming that each node (MA or AS) has no information about arrival rate of classes at other nodes. 

\begin{figure}
\centering
\includegraphics[scale=0.55]{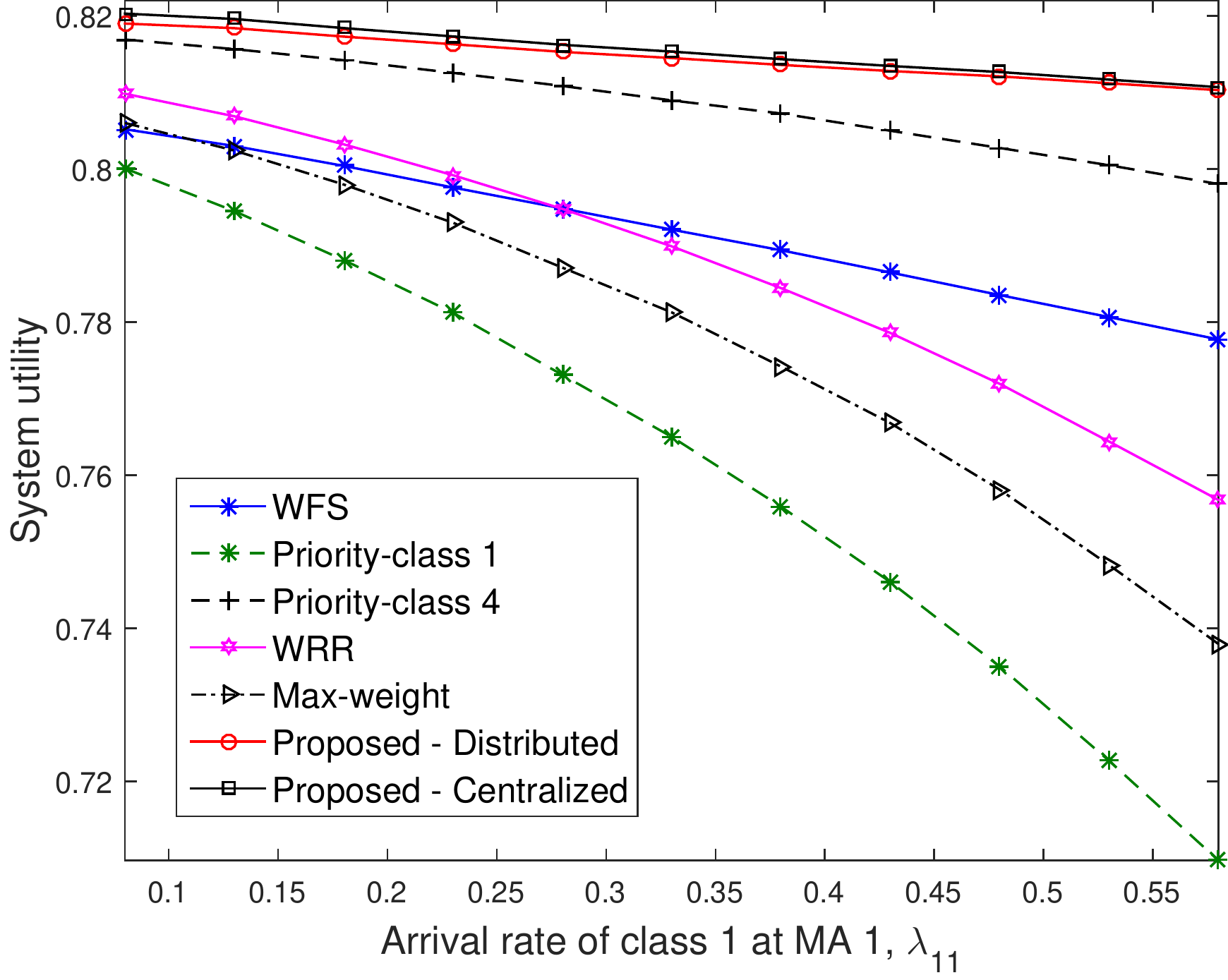}%
\caption{System utility for joint schedulers with $[M, R]=[4, 4]$.}
\label{jointSysUtil} 
\end{figure}

We now extend the previously discussed state-of-the-art schedulers to the joint scheduling problem, by simply reapplying them at each MA and AS. Fig.~\ref{jointSysUtil} compares the system utility of various joint-scheduling schemes as $\lambda_{11}$ is varied while other parameters are kept constant. We note that the centralized optimization scheduler has superior delay-performance as compared to other schedulers and the distributed optimization scheduler results in a near-optimal delay-performance. The loss in optimality is solely due to the Markovian arrival process assumption at AS.

\subsection{Joint Subcarrier Assignment and Packet Scheduling}
In this section, we evaluate the performance of proposed joint subcarrier assignment and packet scheduling scheme. We consider smart metering (SM) as a representative M2M application scenario and set the parameters in our model in accordance with the UCA OpenSG smart grid networks system requirements specification \cite{openSG}. 

We consider that the uplink SM traffic from $4$ different regions in a city is aggregated as specified in Table~\ref{M2Maggregator}.
\begin{table}
\caption{List of M2M aggregators serving the different regions.}
\begin{center}
  \begin{tabular}{ | c | c | c |}
    \hline
    MA Id & Region Served & Number of SMs \\ \hline
    1 & Large industrial plant & 1700 \\ \hline 
    2 & Sparse residential area & 100 \\  \hline
    3 & Medium-size residential area & 800 \\ \hline
    4 & $70 \%$ residential and $30 \%$ commercial area & 1000 \\ \hline 
  \end{tabular}    
  \label{M2Maggregator}
\end{center}
\end{table}

The SM traffic is heterogeneous and comprises of the $4$ important use-cases (i.e., $R=4$  QoS classes) as listed in Table~\ref{M2MuseCases} along with payload size and latency requirements (utility parameters $a_i$ and $b_i$) and arrival rate per device \cite{Nielsen15}. We assume that a certain fraction of the SMs in each region are enhanced Smart Meters (eSM) that provide detailed and frequent measurement reports (Class $1$ in Table~\ref{M2MuseCases}) of the power quality parameters necessary for real-time monitoring and state estimation \cite{Nielsen15}. The exact fraction of eSM traffic is a design parameter and depends on several factors. However, for simplicity, we assume that eSM constitutes $3~\%$ of overall traffic for MA $1$, $3$ and $4$.
\begin{table}
     \caption{List of important M2M use-cases and their characteristics.}
\begin{center}
  \begin{tabular}{ | c | c | c | c | c |}
    \hline
    M2M Use Case & Payload & $(a,b,\beta)$ & Arrivals/sec/SM \\ \hline
    eSM reporting & 2500 B& (4,1,0.95) & $1$ \\ \hline 
    Real-time pricing & 25 B & (1.5,5,0.85) & $1.1*10^{-3}$\\ \hline
    Meter reading (on-demand) & 2400 B & (0.8,5,0.65) & $2.89*10^{-7}$ \\ \hline 
    Regular meter readings & 1560 B & (0.05,70,0.35) & $2.78*10^{-4}$(com.)/ $6.94*10^{-5}$(res.) \\ \hline
  \end{tabular}
  \label{M2MuseCases}
\end{center}
\end{table}

The subcarrier bandwidth, $W$, is set to $15$~kHz and the received SNR at the AS under AWGN condition is $\eta=10$~dB. We assume block i.i.d. Rayleigh fading for each MA-AS channel link. We repeat the proposed procedure for the optimal subcarrier allocation and packet scheduler capacity each time the channel changes significantly between any MA-AS link. For our simulation, we have $[{\|h_1\|}^2, {\|h_2\|}^2, {\|h_3\|}^2, {\|h_4\|}^2]=[3.205, 0.6454, 1.12, 1.778]$ and thus subcarrier capacity as $[C_{s,1}, C_{s,2}, C_{s,3}, C_{s,4}]=[75.66, 43.47, 54.13, 63.47]$~kbps. The AS service rate is set to $C_o=4$~Mbps. The total number of subcarriers is $N=40$.

We solve the centralized MINLP problem using the standard branch and bound method for solving integer optimization problem. We also determine an upper bound on the optimal solution by relaxing the integer constraints on the subcarrier allocation to allow for real-valued subcarrier allocation. We evaluate the performance of our proposed distributed optimization framework against these two results.

\begin{figure}
\centering
\includegraphics[scale=0.45]{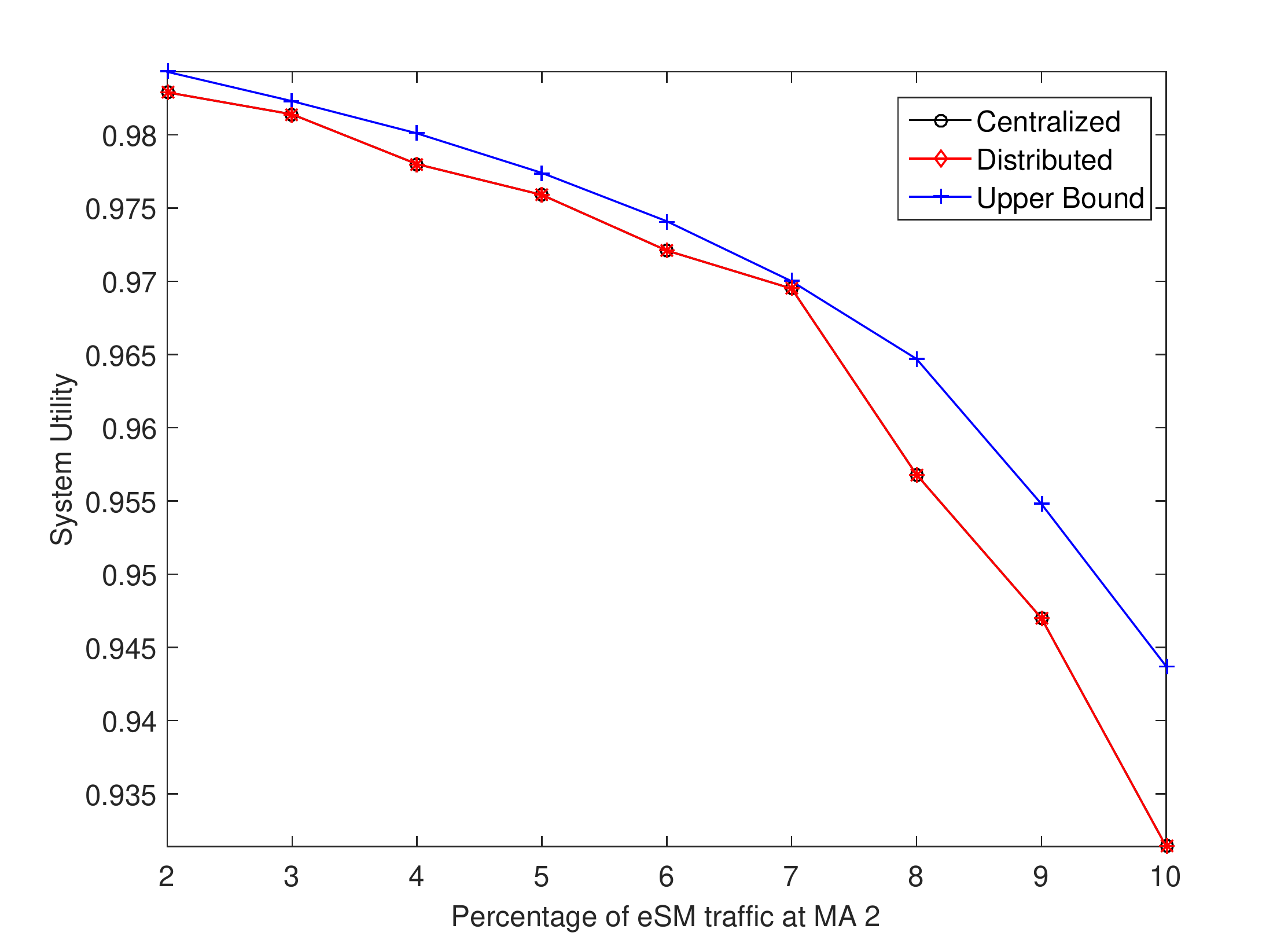}%
\caption{System utility for joint subcarrier allocation and packet scheduling at MAs and AS. The upper bound on the optimal solution is obtained by allowing for real-valued subcarrier allocation.}
\label{jointSysUtil_ChanPacket} 
\end{figure}

Fig.~\ref{jointSysUtil_ChanPacket} shows the plot of system utility as the eSM traffic at MA $2$ is increased from $2\%$ to $10\%$. We note that there is no loss in optimality by resorting to the proposed distributed optimization framework, yet it significantly reduces the complexity of the original NP-hard problem. Furthermore, the system utility for both centralized and distributed schemes is fairly close to the upper-bound obtained by relaxing the optimization problem to permit real-valued subcarrier allocation. 

\begin{figure}
\centering
\includegraphics[scale=0.45]{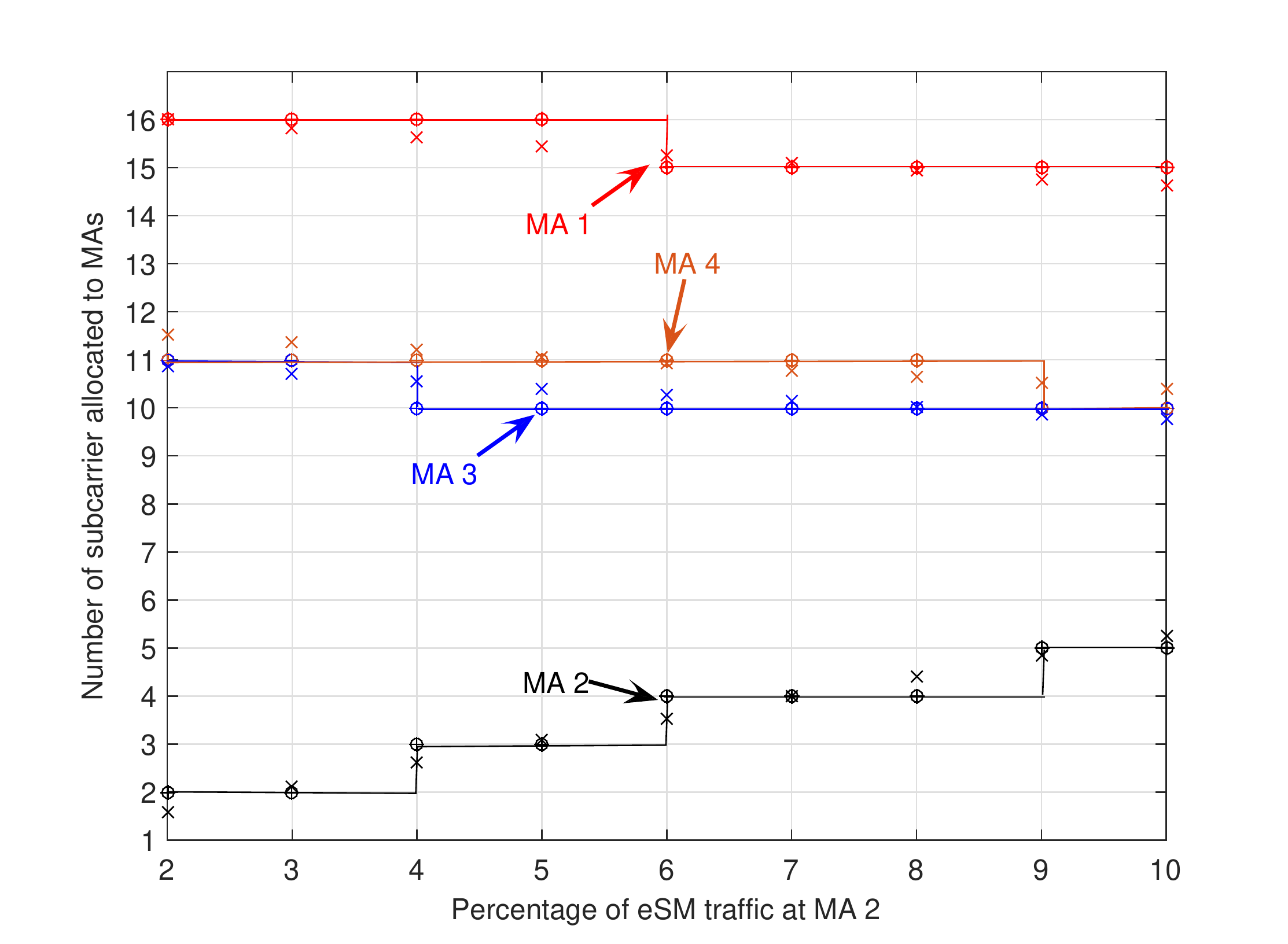}%
\caption{Subcarrier allocation to MAs as eSM traffic at MA $2$ is increased. We plot the results for the following: (a) centralized integer subcarrier allocation \lq o\rq, (b) real-valued subcarrier allocation \lq x\rq, (c) distributed integer subcarrier allocation \lq +\rq.}
\label{subcarrierAllocation} 
\vspace{15pt}
\end{figure}

Fig.~\ref{subcarrierAllocation} illustrates the dynamics of subcarrier allocation to MAs as the eSM traffic at MA $2$ is increased from $2\%$ to $10\%$. We first note that MA $2$ is assigned the least number of subcarriers because it provides service to a sparse residential area with just $100$ SMs. As eSM traffic at MA $2$ is increased, more subcarriers are assigned to MA 2 due to real-time, low latency requirements and large payload of eSM traffic. Due to this the number of subcarriers allocated to other MAs starts to decrease. Since MA $3$ has the least number of SMs among MA $1$, $3$ and $4$, it is the first one to get hit with subcarrier reduction. This is followed by MA $1$ and lastly MA $4$. Lastly, we note that the centralized and proposed distributed subcarrier allocation results in same subcarrier assignment and is fairly close to the real-valued subcarrier allocation.

\begin{figure}
\centering
\includegraphics[scale=0.45]{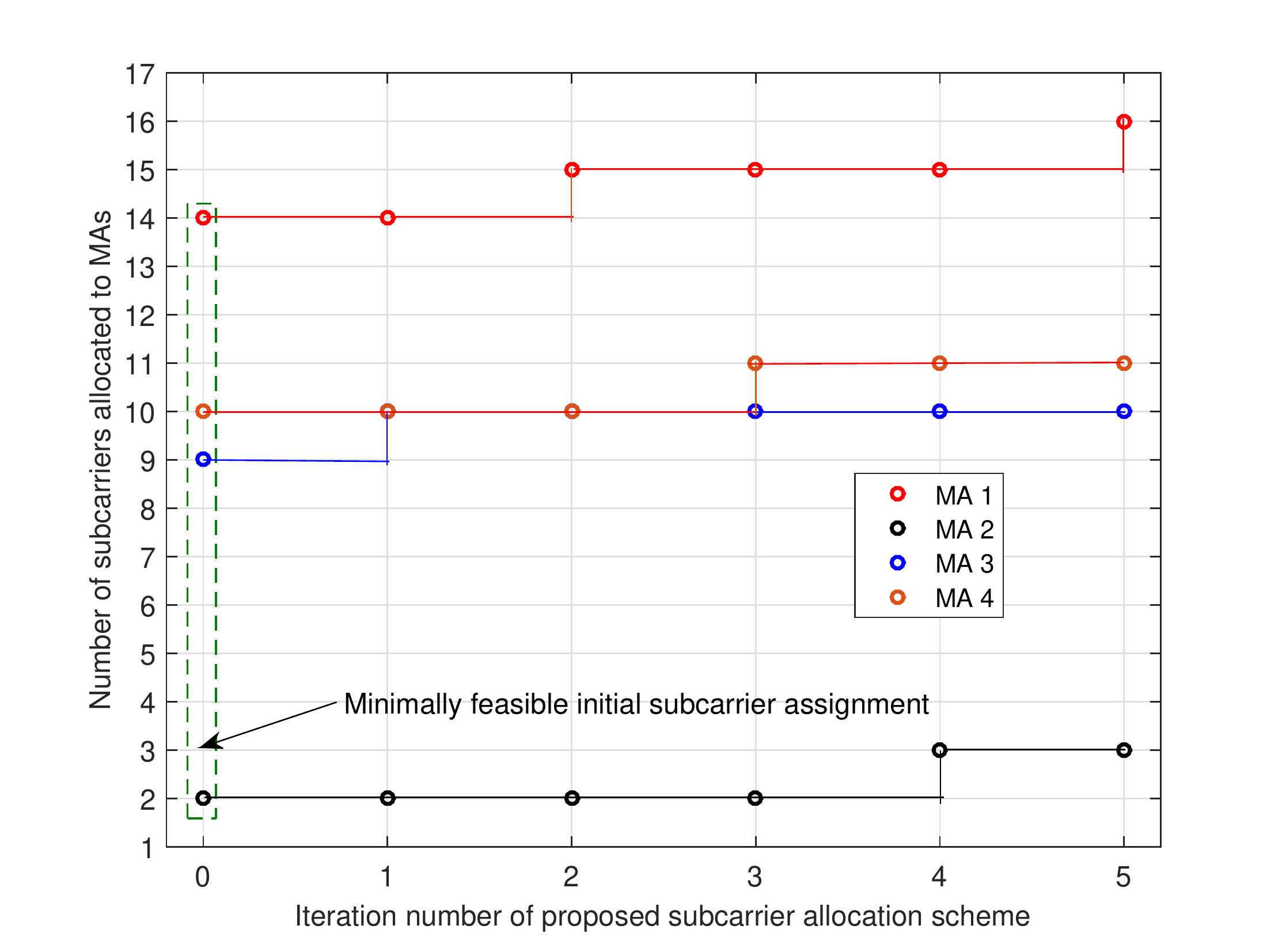}%
\caption{Iterative procedure of distributed subcarrier allocation to MAs when MA $2$ has $4\%$ eSM traffic. The eSM traffic at MA 2 is kept constant at $4\%$. Iteration $0$ corresponds to the minimally feasible assignment used to initialize Algorithm~\ref{propChannelAss}. }
\label{subcarrierAlloc_iteration} 
\vspace{15pt}
\end{figure}

Fig.~\ref{subcarrierAlloc_iteration} illustrates the subcarrier allocation at end of each iteration in the proposed distributed optimization framework.  The iteration $0$ corresponds to the minimally feasible assignment used to initialize Algorithm~\ref{propChannelAss}. We note the subcarrier assigned to MA 2 (which has least amount of M2M traffic) remains fairly constant at $2$ except towards the end of algorithm when it is assigned $3$ subcarriers.

\section{Conclusions}
\label{concl}
In this paper, we studied delay-optimal packet scheduling strategies for a M2M uplink with heterogeneous data arriving at M2M Application Server (AS) via multiple M2M Aggregators (MAs). We classified the uplink traffic into multiple QoS classes based on the M2M data characteristics such as maximum delay budget, packet size and arrival rate requirements. We used sigmoidal function to represent the service utility of a class as a function of packet delay.  We exploited the fact that the average class delay for any work-conserving scheduling policy can be realized by appropriately time-sharing between all preemptive priority scheduling policies. So we determine the delay-optimal packet scheduler at AS by solving for the optimal fraction of time-sharing between all priority scheduling policies so as to maximize a proportionally-fair system utility metric. The computational complexity of the optimal scheduler is significantly reduced by iteratively solving multiple, small optimization problems rather than a single big optimization problem. 

We then extend our work to joint MA-AS channel assignment and packet scheduling strategies at the MAs and AS. We considered both centralized single-stage optimization at AS and then a distributed iterative optimization framework that lets each MA and AS to solve for its locally optimal scheduler while fixing the local scheduler at other nodes. We showed that the  distributed optimization has low computational complexity, with minimal information exchange overhead and converges quickly to the centralized optimization result.  

 Using Monte-Carlo simulations, we verified the optimality of the iterative scheduler and showed that it outperforms various state-of-the-art schedulers (such as WRR, WFS, max-weight and priority scheduler) in terms of system utility. It also results in near-minimal delay jitter for delay-sensitive traffic at the expense of higher delay jitter for delay-tolerant traffic. Lastly, we again note the proposed joint MA-AS packet scheduler outperforms other scheduling policies in terms of delay-performance.
\bibliographystyle{ieeetr}	

\bibliography{referencesJournal}	

\end{document}